\documentclass[twocolumn,amsmath,amssymb,floatfix,prl,nobibnotes,superscriptaddress]{revtex4-1}

\usepackage{amsmath,amssymb,natbib,bm,graphicx,url,epsf,color}
\usepackage[english]{babel}
\usepackage{wasysym}
\usepackage{MnSymbol}
\usepackage{amsmath}
\usepackage{graphicx}
\usepackage{multirow}
\usepackage{ulem}
\usepackage[colorinlistoftodos]{todonotes}
\usepackage[colorlinks=true, allcolors=blue]{hyperref}

\usepackage{pdfpages}
\usepackage[labelfont=bf,justification=raggedright,labelsep = space]{caption}

\newcommand{\VD}{V_{\mathrm{D}}}
\newcommand{\Vd}{V_{\mathrm{D}}}
\newcommand{\VS}{V_{\mathrm{S}}}
\newcommand{\Vs}{V_{\mathrm{S}}}

\newcommand{\VPa}{V_{\mathrm{P1}}}
\newcommand{\VPb}{V_{\mathrm{P2}}}

\newcommand{\dIdVpb}{\partial I_{2}/\partial V_{\mathrm{P}2}}
\newcommand{\Vlength}{V_{\mathrm{L}}}
\newcommand{\Edecay}{E_{\mathrm{d}}}

\makeatletter
\AtBeginDocument{\let\LS@rot\@undefined}
\makeatother

\begin{document}
\title{Relaxation and revival of quasiparticles injected in an interacting quantum Hall liquid}

\author{R. H. Rodriguez}
\affiliation{Universit\'e Paris-Saclay, CEA, CNRS, SPEC, 91191, Gif-sur-Yvette, France
}

\author{F. D. Parmentier}\email[Corresponding author:]{ francois.parmentier@cea.fr}
\affiliation{Universit\'e Paris-Saclay, CEA, CNRS, SPEC, 91191, Gif-sur-Yvette, France
}

\author{D. Ferraro}
\affiliation{
Dipartimento di Fisica, Universit\`a di Genova, Via Dodecaneso 33, 16146, Genova, Italy
}
\affiliation{
SPIN-CNR, Via Dodecaneso 33, 16146, Genova, Italy
}
\author{P. Roulleau}
\affiliation{Universit\'e Paris-Saclay, CEA, CNRS, SPEC, 91191, Gif-sur-Yvette, France
}
\author{U. Gennser}
\affiliation{Universit\'e Paris-Saclay, CNRS, Centre de Nanosciences et de Nanotechnologies (C2N), 91120 Palaiseau, France
}
\author{A. Cavanna}
\affiliation{Universit\'e Paris-Saclay, CNRS, Centre de Nanosciences et de Nanotechnologies (C2N), 91120 Palaiseau, France
}

\author{M. Sassetti}
\affiliation{
Dipartimento di Fisica, Universit\`a di Genova, Via Dodecaneso 33, 16146, Genova, Italy
}
\affiliation{
SPIN-CNR, Via Dodecaneso 33, 16146, Genova, Italy
}

\author{F. Portier}
\affiliation{Universit\'e Paris-Saclay, CEA, CNRS, SPEC, 91191, Gif-sur-Yvette, France
}

\author{D. Mailly}
\affiliation{Universit\'e Paris-Saclay, CNRS, Centre de Nanosciences et de Nanotechnologies (C2N), 91120 Palaiseau, France
}

\author{P. Roche}
\affiliation{Universit\'e Paris-Saclay, CEA, CNRS, SPEC, 91191, Gif-sur-Yvette, France
}

\date{\today}

\maketitle

\textbf{The one-dimensional, chiral edge channels of the quantum Hall effect are a promising platform in which to implement electron quantum optics experiments; however, Coulomb interactions between edge channels are a major source of decoherence and energy relaxation. It is therefore of large interest to understand the range and limitations of the simple quantum electron optics picture. Here we confirm experimentally for the first time the predicted relaxation and revival of electrons injected at finite energy into an edge channel. The observed decay of the injected electrons is reproduced theoretically within a Tomonaga-Luttinger liquid framework, including an important dissipation towards external degrees of freedom. This gives us a quantitative empirical understanding of the strength of the interaction and the dissipation.}

\section{Introduction}
 
Electron quantum optics \cite{Bocquillon2014, Bauerle2018} is based on the profound analogy between the transport of single quasiparticles in a quantum coherent conductor, and the propagation of single photons in a quantum optics setup. This has led to seminal electron interferometry experiments realized in edge channels (ECs) of the quantum Hall effect, whether in a Mach-Zehnder geometry \cite{Ji2003} or, recently, in a Hong-Ou-Mandel setup \cite{Bocquillon2013a} where two single-charge excitations emitted at a well-defined energy collide on a quantum point contact, probing their indistinguishable nature. The majority of these experiments have been performed at filling factor $\nu=2$ of the quantum Hall regime, where, for a given carrier density, the quantum Hall effect is the most stable. However, interactions between the two ECs of $\nu=2$ have been shown to lead to decoherence as well as energy relaxation. The latter corresponds to the fact that energy can be transferred from one EC to the next, even in absence of tunneling between the two. This strongly challenges the simple picture of electron quantum optics, and raises the crucial question of the nature of the excitations that actually are interfering in the aforementioned experiments.

The first investigations of decoherence and energy relaxation at $\nu=2$ involved biased quantum point contacts to generate a broadband, out-of-equilibrium distribution function that was probed using Mach-Zehnder interferometry \cite{Neder2006, Roulleau2007, Litvin2008, Roulleau2008, Roulleau2008a} and energy spectroscopy \cite{Altimiras2009, leSueur2010} techniques. From these works emerged a clearer picture of the role of interactions between copropagating ECs, which is well accounted for by a powerful theoretical description in terms of Tomonaga-Luttinger liquid (TLL) physics. In the so-called TLL model, interactions lead to new eigenstates of the system, which are not Fermionic, but charge- and dipole- (or spin-) like plasmons shared by the two edge channels \cite{Sukhorukov2007, Levkivskyi2008, Degiovanni2009, Degiovanni2010}. The decomposition of a Fermionic excitation in one EC onto the plasmon modes shared by the two ECs gives rise to energy relaxation and decoherence \cite{Degiovanni2009, Degiovanni2010, Ferraro2014}. This model describes particularly well Hong-Ou-Mandel collision experiments using single excitations emitted at finite energy \cite{Bocquillon2013a, Freulon2015, Marguerite2016, Marguerite2017}.

Underlying the TLL model is the assumption that the interaction between the two ECs dwarfs all other energies. This means that although the energy of a carrier injected into one of the ECs will be redistributed between the two interacting ECs, the system will conserve its total energy. How valid this assumption is remains an important question, as a number of the basic predicted features of the evolution of a quasiparticle emitted at finite energy remain to be confirmed experimentally. The shape of the energy distribution of finite-energy quasiparticles, which is referred to as the quasiparticle peak, has so far not been observed in the quantum Hall regime; nor has its evolution during propagation. 

In fact, probing the quasiparticle peak is of crucial importance, since it would directly reflect the wavepackets of single particles that are manipulated in quantum optics, and its behavior could establish unambiguously characteristics specific to the TLL model. One such potential feature is the remarkable ability to partially regenerate the initial excitation \cite{Ferraro2014}. This is analogous to Rabi oscillations, where a system oscillates between two states that are not proper eigenstates due to their mutual interaction. Specifically, the TLL predicted regeneration of an initial excitation comes about through the 'catching up' and recombination of a fast-propagating charge plasmon with a slower dipole plasmon (animations illustrating the effect can be found in the supplemental material of ref.~\cite{Ferraro2014}). However, this resurgence has only been indirectly observed in Mach-Zehnder interferometry experiments with biased quantum point contacts \cite{Neder2006, Litvin2008, Gurman2016}, whereas it should clearly appear as a revival of the quasiparticle peak at finite length and energy \cite{Ferraro2014}.
Furthermore, recent experiments using such finite energy quasiparticles in other schemes revealed important qualitative inconsistencies with the TLL model. First, spectroscopy experiments showed that a sizable portion of the energy injected in the system was lost to additional degrees of freedom, not included in the TLL model \cite{leSueur2010}. Second, finite energy excitations were shown to interfere within a Mach-Zehnder setup with a visibility that decreased, but remained finite even at high energy instead of fully vanishing as predicted \cite{Tewari2016}. Very recently, an experiment using an energy spectroscopy technique similar to the one reported in the present paper showed that quasiparticles can exchange energy between spatially distinct parts of the circuit \cite{Krahenmann2019, Fischer2019}. While this result can explain the missing energy reported in \cite{leSueur2010}, it is again in contrast with the TLL model. This series of inconsistencies raises a crucial question: is there merely a missing ingredient in the TLL model for it to fully describe the physics of interacting edge channels, or is it necessary to replace it with a different theory?
Indeed, a recent competing theoretical description \cite{Lunde2016} is qualitatively compatible with the early energy spectroscopy experiments \cite{Altimiras2009, leSueur2010}. Based on a Fermi liquid description of the edge channels, and the assumption that electron-electron interactions do not conserve momentum, this model predicts that the quasiparticle peak gradually broadens and shifts towards lower energies while both edge channels are warmed up. Contrary to double step distribution functions obtained with a biased quantum point contact, which yield similar results within both models, the predicted behavior of finite energy quasiparticles is thus strikingly different, as the TLL model predicts the quasiparticle peak to diminish in amplitude, and then to revive, while its position and width remain constant.

To answer the above question, we have performed an experimental investigation of the energy relaxation of energy-resolved quasiparticles, showing a clear observation of the quasiparticle peak at $\nu=2$. We show that while the quasiparticle peak is strongly suppressed with the injection energy and the propagation length, it clearly undergoes a revival at intermediate energy and length before disappearing into a long-lived state that is not fully thermalized. The observed evolution of the quasiparticle peak allows us to unambiguously discriminate between the two models. We show that the TLL model can be refined in order to explain our results by including dissipation towards external degrees of freedom, and, by spatially separating the two edge channels with an additional gate, we unambiguously demonstrate the role of edge channel coupling.

\section{Results}

\subsection{Experimental approach}

\begin{figure}[h!]
\centering
\includegraphics[width=0.45\textwidth]{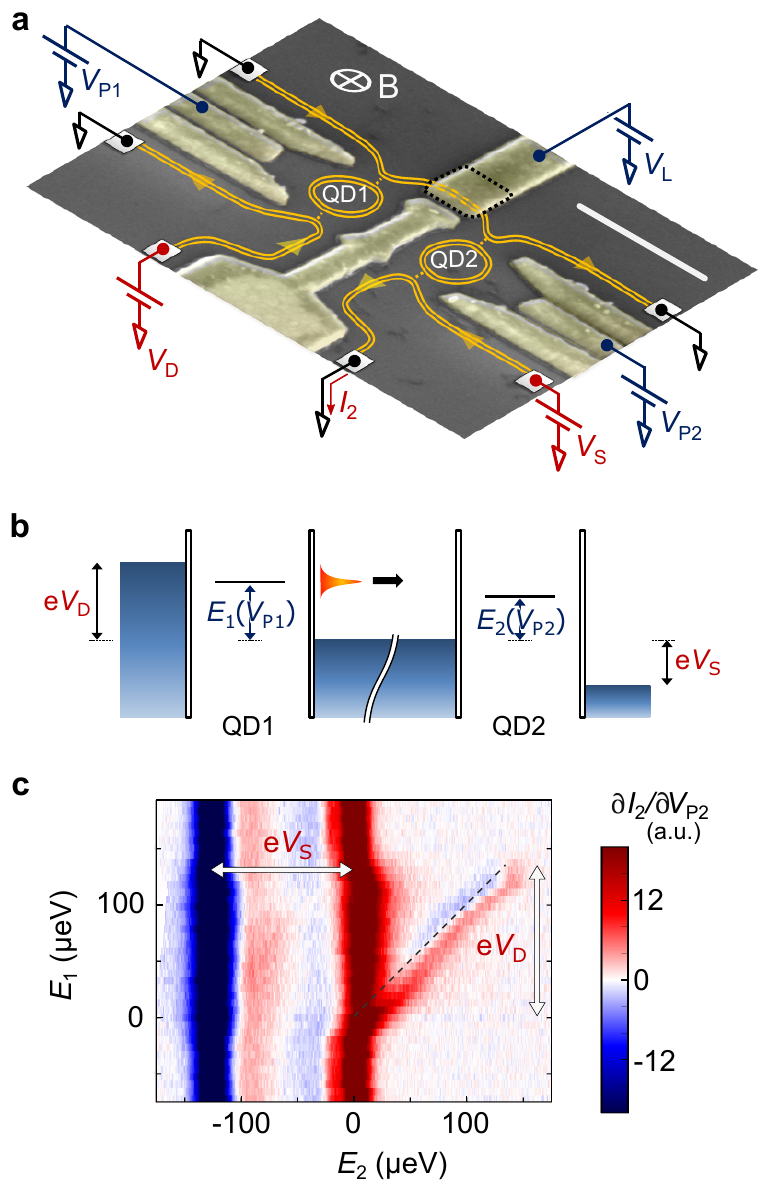}
\caption{\label{fig1}\textbf{$\vert$ Principle and implementation of the experiment. a,} False-colour scanning electron micrograph of a typical sample. The ECs at $\nu=2$ are depicted in orange. The large ohmic contacts located away from the center of the sample are depicted by the gray squares. The white scale bar corresponds to $500~$nm. \textbf{b,} Energy-scale sketch of the experiment. The two QDs are depicted by a single resonance at energy $E_1(V_{\mathrm{P}1})$ and $E_2(V_{\mathrm{P}2})$, respectively. The emitted quasiparticles are depicted by the orange bell-shaped curve. \textbf{c,} Raw transconductance $\dIdVpb$ of the second QD measured as function of $E_2(V_{\mathrm{P}2})$ (x-axis) and $E_1(V_{\mathrm{P}1})$ (y-axis). The thick vertical (resp. horizontal) arrow indicates the span of the drain (resp. source) potential $\VD$ (resp. $\VS$). The $y=x$ dashed line is a guide for the eye.}
\end{figure}

We have followed the approach proposed in \cite{Lunde2016, Takei2010}, and recently applied in \cite{Krahenmann2019}, in which one injects quasiparticles at a well-defined energy into an edge channel using a first quantum dot (QD) in the sequential tunneling regime. The injected quasiparticles then propagate over a finite length $L$, after which we perform a spectroscopy of the energy distribution function $f(E)$ of the quasiparticles using a second downstream quantum dot as energy filter. This spectroscopy technique combined with a quantum point contact to generate excitations was previously used in \cite{Altimiras2009, leSueur2010, Altimiras2010, Itoh2018}. A very similar setup was used to investigate charge transfer processes between distant quantum dots in the absence of a magnetic field \cite{Rossler2014}; furthermore, a recent spectroscopy experiment showed that at vastly higher energies (in the $0.1~$eV range), electrons in an edge channel decay by coupling to optical phonons~\cite{Ota2019}. It is also worth noting that other experimental techniques can be used to probe the energy distribution function, by measuring shot noise \cite{Gabelli2013}, or by performing a quantum tomography of the excitation injected in the edge channel \cite{Grenier2011b, Jullien2014, Bisognin2019, Fletcher2019}. The latter is known for being amongst the most challenging experiments undertaken so far in electron quantum optics.

The devices' geometry is depicted in Fig.~\ref{fig1}a. The two chiral edge channels of $\nu=2$ are depicted as orange lines. The quantum dots are defined electrostatically, and can be independently controlled using the plunger gate voltages $\VPa$ and $\VPb$. Both QDs are tuned to transmit only the outer edge channel. Quasiparticles in the outer edge channel stemming from the drain electrode are thus transmitted across the first dot QD1, and propagate along the outer edge channel connecting the first dot to the second dot QD2. A length gate, controlled by the voltage $\Vlength$, is used to increase the propagation path by diverting the ECs around the square area delimited by black dashed lines in Fig.~\ref{fig1}a (a 200~nm insulating layer of SU-8 resist separates the rest of the gate from the surface of the sample). Several samples have been measured; here we show results obtained on three different devices, with nominal propagation lengths $L=480~$nm, $L=750~$nm, and $3.4~\mu$m. Using the length gate on the first two devices yields the additional lengths $L\approx1.3~\mu$m (long path for the $L=480~$nm device) and $L\approx2.17~\mu$m (long path for the $L=750~$nm device - see Supplementary Information Note 1 for details on the devices, including the estimation of the lengths). Fig.~\ref{fig1}b depicts the energy configuration of the two dots: a negative voltage $\Vd$ is applied to the drain contact while the contacts connected to the edge channels flowing between the two dots are grounded, defining the zero of energy in our experiment. A narrow single resonance of QD1 is tuned inside the transport window at an energy $E_1(\VPa)$, defining the quasiparticle injection energy. We measure the transconductance $\dIdVpb$ of QD2 while sweeping the energy $E_2(\VPb)$ of a narrow single resonance in this dot that defines the detection energy. A calibration of both QDs is performed to extract their respective lever arms, linking the plunger gates voltages $V_\mathrm{Pi}$ to the energies $E_i$ (see Supplementary Note 2). This allows us, after compensating for the small crosstalks between the two plunger gates, to directly probe the dependence of $\dIdVpb$ with the detection energy $E_2$ for different values of the injection energy $E_1$. This signal is proportional to $-\partial (\Delta f(E))/\partial E$, where $\Delta f(E)=f(E)-f_\mathrm{S}(E)$ is the difference of the energy distribution functions on either side of QD2 \cite{Altimiras2009, leSueur2010, Altimiras2010, Itoh2018}, convoluted with the lineshape of the resonance of QD2 ($f_\mathrm{S}(E)$ is the distribution function of the source EC). This convolution mostly affects the width of the features in the transconductance (see Supplementary Note 3). In the following, all widths discussed are convoluted widths. We separate the two contributions of $f(E)$ and $f_\mathrm{S}(E)$ by applying a positive voltage $\Vs$ to the source contact. This is illustrated in Fig.~\ref{fig1}c, which shows a typical measurement of $\dIdVpb$ as a function of $E_1$ and $E_2$, for $L=480~$nm. The source and drain potentials, shown as thick arrows in Fig.~\ref{fig1}c, are set to $\mathrm{e}\VD \approx - \mathrm{e}\Vs \approx 125~\mu$eV, with e$\approx-1.6\times10^{-19}~$C the electron charge. The three main features appearing on this map are \textit{i)} the blue (negative) vertical line at $E_2=\mathrm{e}\Vs\approx -125~\mu$eV, corresponding to $\partial f_\mathrm{S}(E)/\partial E$, \textit{ii)} the red (positive) vertical line at $E_2\approx0~\mu$eV, corresponding to the low-energy part of $\partial f(E)/\partial E$, and \textit{iii)} the oblique line following a $y=x$ line (black dashed line), corresponding to the emitted quasiparticles which are detected after their propagation. Note that no signature of Auger-like processes \cite{Krahenmann2019} (which would appear as diagonal lines dispersing in a direction opposite to the black dashed line) has been identified in any of the transconductance maps we obtained. We integrate the transconductance so as to obtain the energy distribution function $f(E)$, which we discuss in the rest of this paper. 

\subsection{Measured distribution functions}

\begin{figure}[h]
\centering
\includegraphics[width=0.42\textwidth]{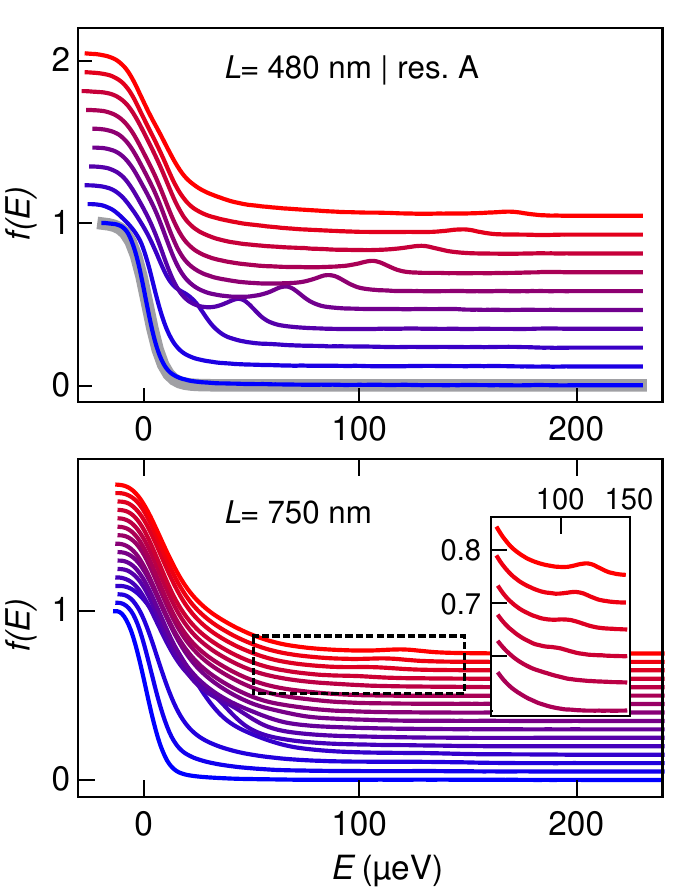}
\caption{\label{fig2}\textbf{$\vert$ Measured distribution functions.} Top panel: measured $f(E)$ for $L=480~$nm. Each curve, offset for clarity, corresponds to an increment of the injection energy $\delta E_1\approx 21~\mu$eV, from $E_1=-21~\mu$eV (blue) where no additional quasiparticles are emitted, to $E_1=173~\mu$eV (red). The thick gray line is a Fermi function fit of the data at $E_1=-21~\mu$eV. Bottom panel: measured $f(E)$ for $L=750~$nm. Each curve corresponds to a increment of the injection energy $\delta E_1\approx 9~\mu$eV, up to $E_1=121~\mu$eV (red). The inset is a zoom on the region delimited by the dashed-line square. In all panels, the vertical offset is equal to $5.5\times10^{3}\delta E_1$.
}
\end{figure}

Fig.~\ref{fig2} shows measurements of $f(E)$ for $L=480~$nm (top panel) and $L=750~$nm (bottom panel). The injection energy $E_1$ is gradually increased from negative values (blue curves), where the resonance of QD1 is outside the bias window, to large positive values $E_1 >100~\mu$eV (red curves), where we expect to detect quasiparticles at high energy. The measured $f(E)$ curves evolve from a Fermi function at low temperature (the apparent temperature is increased to $\sim 40~$mK by the convolution with the resonance of QD2, see SUpplementary Note 3) to strongly out-of-equilibrium distribution functions showing a distinct quasiparticle peak at finite energy. This is particularly striking for the shortest distance (top panel), where the peak clearly appears even at the largest energy $E_1=173~\mu$eV (note that the peak was not observed in ref.~\cite{Krahenmann2019}, where the propagation length was $\sim1.5~\mu$m). The peak position increases linearly with $E_1$, while its amplitude decreases. In contrast, for a path only $50~\%$ longer, the peak amplitude is strongly suppressed; however, after vanishing at $E_1\sim90~\mu$eV, it reappears as $E_1$ is further increased (see inset in the bottom panel of Fig.~\ref{fig2}). The clear presence of a quasiparticle peak, its strong decay, and its subsequent revival at intermediate lengths are consistently observed in our experiment, and are the main results of this paper. In the following, we quantitatively analyze the measured $f(E)$, and compare our results with the leading theories.

\subsection{Quasiparticle peak analysis}

\begin{figure*}[ht!]
\centering
\includegraphics[width=0.97\textwidth]{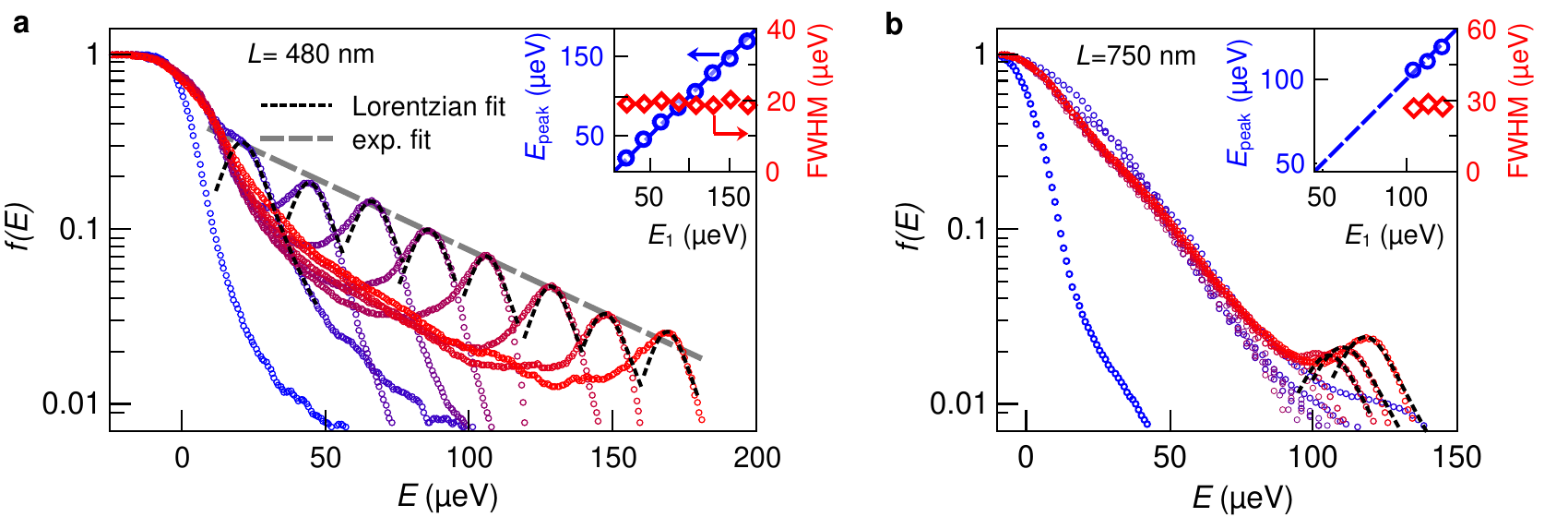}
\caption{\label{fig3} \textbf{$\vert$ Exponential decay and revival of the quasiparticle peak.} Measured $f(E)$ for $L=480~$nm (\textbf{a}) and $L=750~$nm (\textbf{b}), in semi-log scale. The circles are experimental data, with the same dataset as in Fig.~\ref{fig2}, ranging from $E_1={-21, -17}~\mu$eV (blue), to $E_1={173, 121}~\mu$eV (red). The black dashed lines are Lorentzian fits of the quasiparticle peak, and the grey dashed line in \textbf{a} is an exponential fit of the maxima of the Lorentzian fits, with a characteristic energy $\Edecay\approx56~\mu$eV. Insets: Center $E_{\mathrm{peak}}$ (blue circles, left Y-axis) and full width at half maximum FWHM (red diamonds, right Y-axis) of the Lorentzian fits, plotted versus injection energy $E_1$. The size of the symbols indicate our experimental accuracy. The blue dashed line in the insets is a $y=x$ line. 
}
\end{figure*}

Fig.~\ref{fig3} shows a semi-log scale plot of the data shown in Fig.~\ref{fig2}, illustrating our analysis. For $L=480~$nm (Fig.~\ref{fig3}a), the quasiparticle peak is well fitted by a Lorentzian peak without any offset, shown as dashed black lines. Remarkably, the energy position $E_\mathrm{peak}$ of the peak matches the injection energy $E_1$, and its full width at half maximum (FWHM) remains constant as $E_1$ is increased (see inset in Fig.~\ref{fig3}a). This observation, which was consistent in all data where the quasiparticle peak is distinguishable, is in direct contradiction with the predictions of \cite{Lunde2016}, but in agreement with the TLL model. Furthermore, the semi-log scale shows that the maximum of the quasiparticle peak follows an exponential decay (gray dashed line) over more than an order of magnitude. For $L=750~$nm (Fig.~\ref{fig3}b), the peak is strongly suppressed. However, while the peak only shows up as a faint bump at low $E_1$ and has vanished for intermediate $E_1$, it appears clearly at large $E_1$, and can again be fitted by a Lorentzian with preserved width and position. In addition, the peak height increases with the injection energy, as seen in Fig.~\ref{fig2}. We observed the revival in several realizations of the experiment in the same $L=750~$nm device, with different gating conditions, and during different cooldowns (note that while we did not observe the revival for $L=480~$nm, we show below that it is expected to occur at significantly higher $E_1$, outside our spectroscopy range - see methods). While those observations clearly are characteristic features of the TLL model, it is not the case for the apparent exponential decay of the peak at $L=480~$nm. Another discrepancy is the fact that the low energy part of distribution functions, away from the quasiparticle peak, seem to be (at least to some extent) independent of the injection energy, whereas it should become broader with increasing $E_1$. This strongly suggest that dissipation - that is, loss of energy towards other degrees of freedom than the plasmon modes - needs to be taken into account. The presence of dissipation was already identified in previous works \cite{leSueur2010, Bocquillon2013b}, and particularly in \cite{Krahenmann2019} where it manifested as long-distance Auger-like processes.

\subsection{Modelling dissipation in the TLL model}

\begin{figure}[ht!]
\centering
\includegraphics[width=0.42\textwidth]{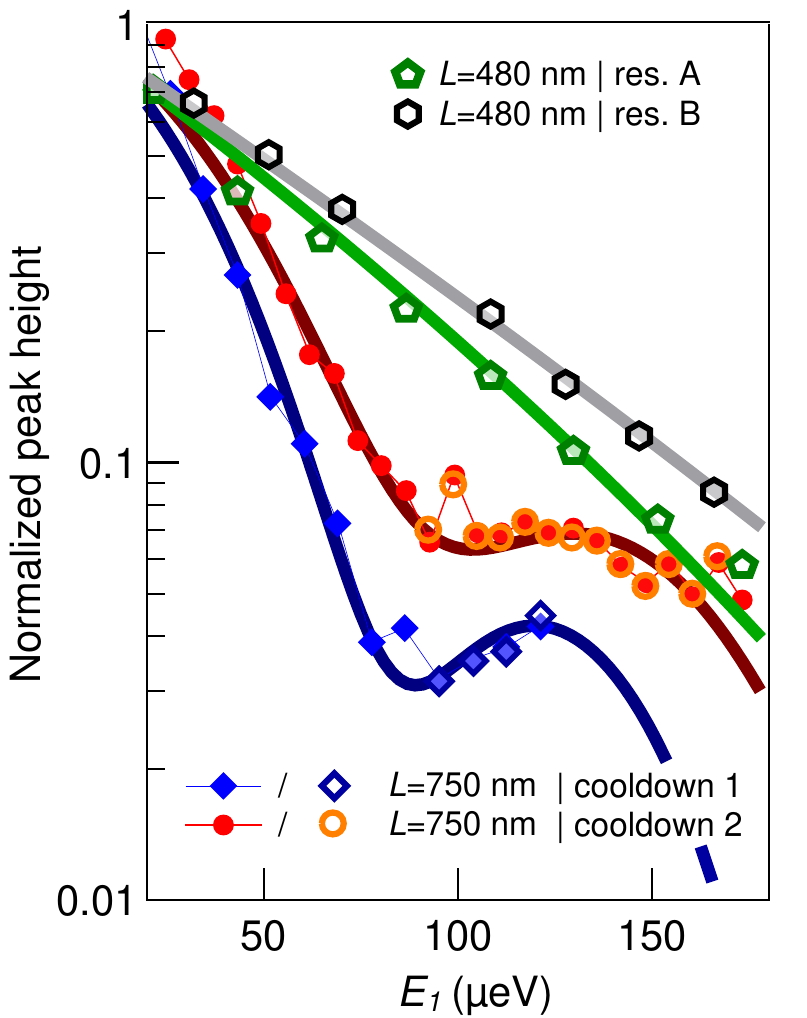}

\caption{\label{fig4} \textbf{$\vert$ TLL fits with dissipation.} Normalized peak height plotted in semi-log scale as a function of injection energy $E_1$ for $L=480$~nm (green pentagons – resonance A, and black hexagons – resonance B) and $L=750$~nm (blue diamonds – cooldown 1, and red circles – cooldown 2). The size of the symbols corresponds to our typical experimental accuracy. The open symbols are the Lorentzian fits heigths, and the full symbols the direct extractions of the amplitude $f(E_1)$ for $L=750$~nm. The thick lines are fits of the data using the TLL model including dissipation described in the text.
}
\end{figure}

\begin{table*}[h!]
\begin{tabular}{ c c c c c c c c c c }
\hline
Sample & $v_2$~(km$~\mathrm{s}^{-1}$) & $v_\rho$~(km$~\mathrm{s}^{-1}$) & $v_\sigma$~(km$~\mathrm{s}^{-1}$) & $\alpha$ & $\theta$ & $\gamma_0$ & $E_0$~($\mu$eV) & $E_\gamma$~($\mu$eV)\\
\hline
480~nm & \multirow{2}{5em}{\centering 48} & \multirow{2}{5em}{\centering 89} & \multirow{2}{5em}{\centering 35} & \multirow{2}{5em}{\centering 1.6} & \multirow{2}{5em}{\centering 0.16~$\pi$} & \multirow{2}{5em}{\centering 0.43} & \multirow{2}{5em}{\centering 403} & \multirow{2}{5em}{\centering 142} \\
res. A &  &  &  & &  & & & \\
\hline
480~nm & \multirow{2}{5em}{\centering 48} & \multirow{2}{5em}{\centering 92} & \multirow{2}{5em}{\centering 42} & \multirow{2}{5em}{\centering 1.8} & \multirow{2}{5em}{\centering 0.11~$\pi$} & \multirow{2}{5em}{\centering 0.43} & \multirow{2}{5em}{\centering 535} & \multirow{2}{5em}{\centering 147}\\
res. B &  &  &  & &  & & & \\
\hline
750~nm & \multirow{2}{5em}{\centering 38} & \multirow{2}{5em}{\centering 101} & \multirow{2}{5em}{\centering 17} & \multirow{2}{5em}{\centering 2.1} & \multirow{2}{5em}{\centering 0.17~$\pi$} & \multirow{2}{5em}{\centering 0.13} & \multirow{2}{5em}{\centering 85} & \multirow{2}{5em}{\centering 342}\\
cooldown 1 &  &  &  & &  & & & \\
\hline 
750~nm & \multirow{2}{5em}{\centering 38} & \multirow{2}{5em}{\centering 118} & \multirow{2}{5em}{\centering 18} & \multirow{2}{5em}{\centering 2.6} & \multirow{2}{5em}{\centering 0.15~$\pi$} & \multirow{2}{5em}{\centering 0.12} & \multirow{2}{5em}{\centering 95} & \multirow{2}{5em}{\centering 452} \\
cooldown 2 &  &  &  & &  & & & \\
\hline 
\end{tabular}
\caption{\label{table1} \textbf{$\vert$ TLL fits parameters.} Lowest Fermi velocity $v_2$, charge and dipole plasmon velocities $v_\rho$ and $v_\sigma$, Fermi velocities ratio $\alpha$, effective inter-EC coupling $\theta$, friction coefficient $\gamma_0$, revival energy $E_0$ and exponential decay characteristic energy $E_\gamma$ extracted from the fits shown in Fig.~\ref{fig4}.
}

\end{table*}

A simple way to include dissipation in the TLL model (see Supplementary Note 8 for details of the model) consists in introducing an \textit{ad hoc} linear friction term in the equations of motion for the bosonic fields describing the charge and dipole plasmon modes \cite{Levkivskyi2008, Degiovanni2009, Ferraro2014, Braggio2012}. Because of interactions, assumed here to be short-ranged, these modes are shared by the two ECs, and their respective velocities $v_\rho$ (charge mode) and $v_\sigma$ (dipole mode) depend on the Fermi velocities $v_1$, $v_2$ in each EC in absence of interactions, as well as on the coupling $u$ between the ECs. These parameters combine into an effective mixing angle $\theta$, defined as $\mathrm{tan}(2\theta)=2u/(v_1-v_2)=2u/v_2(\alpha-1)$, which is zero when the two ECs do not mix, and $\pi/4$ for maximal coupling. This reflects the fact that even if the interaction $u$ is small, the ECs can become maximally coupled if they propagate at exactly the same velocity. In this strong coupling limit, and in absence of dissipation, the quasiparticle peak height is given by a characteristic squared Bessel function $J_0^2(2.5 \times E_1 / E_0)$, with $E_0=5\hbar v_\rho v_\sigma /L(v_\rho -v_\sigma)\approx 5\hbar v_\sigma / L$ \cite{Degiovanni2009, Ferraro2014}. Its oscillatory behavior corresponds to the revival phenomenon, with the first zero occuring at $E_0$. Tuning $\theta$ away from the strong coupling value modifies the Bessel function profile, leading to a lifting up of the zeros. When one includes dissipation, expressions for the quasiparticle peak height are modified, and acquire an exponentially decaying prefactor $\sim\mathrm{exp}(-E_1 /E_\gamma)=\mathrm{exp}(-2\gamma_0 E_1 L /\hbar v_\rho)$, where $\gamma_0$ is the friction coefficient. Note that the model can be further refined by \textit{e.g.} considering non-linear plasmon dispersion \cite{Slobodeniuk2016, Goremykina2018}, or long range interactions \cite{Bocquillon2013b}.

Fig.~\ref{fig4} shows how this model compares to our data at $L=480$ and 750~nm. We plot the extracted Lorentzian peak heights from the 480~nm data shown in Fig.~\ref{fig3}a (green pentagons), as well as for data obtained using a different resonance of QD1 in the same device (black hexagons), versus injection energy $E_1$. Data are normalized by the calibrated transmission of QD1, corresponding to the expected height of the injected peak (see Supplementary Note 1). The exponential decay observed in Fig.~\ref{fig3} is well reproduced by our model (thick green and grey lines). The TLL fits parameters $v_2$, $\alpha$, $\theta$ and $\gamma_0$, as well as corresponding plasmon velocities $v_\rho$, $v_\sigma$ and the characteristic energies $E_0$ and $E_\gamma$ are summed up in Table~\ref{table1}. In particular, the values of the revival energy $E_0$ obtained for the 480~nm sample are much larger than our maximum spectroscopy range $\sim200~\mu$eV, explaining why the revival is not observed in that sample.
We also plot in Fig.~\ref{fig4} the peak height for two different datasets of the $L=750~$nm device. The blue symbols (labeled cooldown 1) correspond to the data shown in Figs.~\ref{fig2} and \ref{fig3}b. The red symbols corresponds to data obtained in a subsequent cooldown of the device, also showing the revival (see SUpplementary Note 5 for additional data and analysis), despite having a different electrostatic environment due to thermal cycling. Importantly, this demonstrates that the observed revival is a robust phenomenon, unlikely to stem from a spurious mesoscopic effect (such as an impurity along the propagation path, or a parasitic resonance in one of the dots). In both datasets, the open symbols correspond to the peak height extracted from the fits at large $E_1$, when the peak becomes visible again. The full symbols correspond to the value $f(E_1)$ of the measured distribution function taken at the injection energy. An important assumption here is that the peak position has not changed relative to the injection energy during propagation, which is validated for both $L$ by the Lorentzian fits. Again, our results are well reproduced by the model including dissipation (thick dark blue and dark red lines), particularly the observed revival, with parameters displayed in Table~\ref{table1}.
Interestingly, because the exponentially decaying prefactor arising from the additional friction term in our model directly depends on the velocity of the charge mode $v_\rho$, we are able to extract all relevant parameters of the TLL model in our experiment. In contrast, the TLL analysis performed on most previous experiments \cite{leSueur2010, Bocquillon2013b, Marguerite2016, Tewari2016, Marguerite2017, Itoh2018} only provided the value of the dipole mode's velocity $v_\sigma$, while implying a strong coupling regime so that $v_\rho\gg v_\sigma$. Using a rather simple refinement of the TLL model, we are thus able to show that, in our experiment, \textit{i)} the Fermi velocities in the two ECs differ typically by a factor $2$, \textit{ii)} the effective EC coupling is moderate, and that \textit{iii)} as a consequence, the difference between the charge and dipole plasmon velocities is not as large as usually assumed. Note that ref.~\cite{Hashisaka2017} demonstrated that those velocities depend on the voltage applied to the gate defining the channel, reporting similar values (up to a factor 2) in our range of gate voltage. We also show that while the friction parameter is highly sample dependent, it does not depend on the QD resonances within a given sample, or on thermal cycling. 

\subsection{Effect of the length gate}

\begin{figure*}[ht]
\centering
\includegraphics[width=0.95\textwidth]{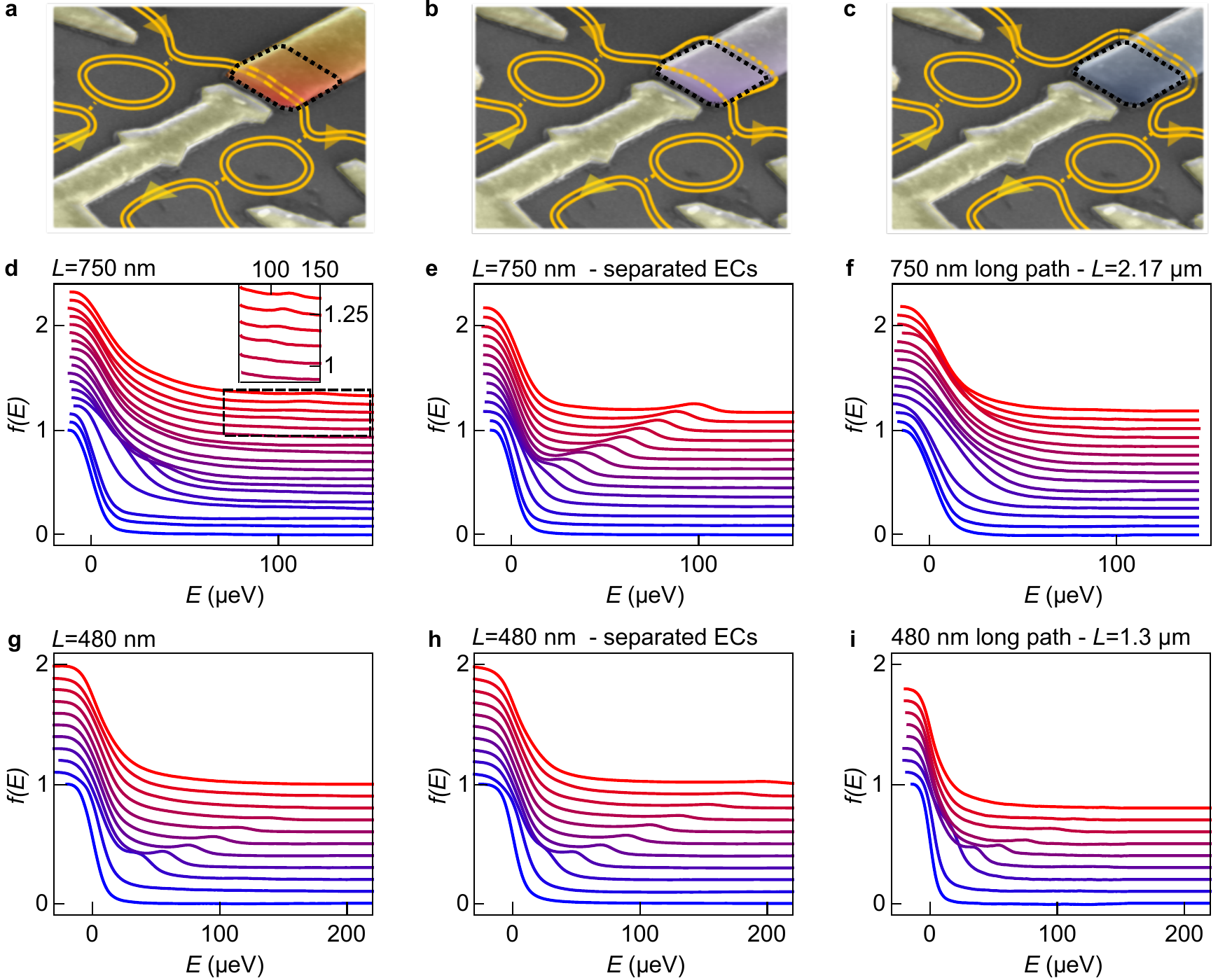}
\caption{\label{fig6} \textbf{$\vert$ Effect of the length gate on energy relaxation. a, b, c,} False-colour scanning electron micrographs of a typical sample, depicting the trajectories of the ECs for $\Vlength=0.2$~V (\textbf{a}, length gate highlighted in red), $\Vlength=-0.1$~V (\textbf{b}, gate highlighted in violet), and $\Vlength=-0.5$~V (\textbf{c}, gate highlighted in dark blue). In \textbf{a}, the ECs copropagate along the short path. In \textbf{b}, the ECs are spatially separated (orange dotted lines) as they flow below the length gate. In \textbf{c}, the ECs copropagate along the long path.
\textbf{d, e, f,} Measured $f(E)$ in the 750~nm device for the configurations depicted in resp. \textbf{a}, \textbf{b}, and \textbf{c}. Each curve, offset for clarity, corresponds to an injection energy increment $\delta E_1\approx 9~\mu$eV, from $E_1\approx -26~\mu$eV (blue) to $E_1\approx 122~\mu$eV (red, \textbf{d}) / $E_1\approx 98~\mu$eV (red, \textbf{e} and \textbf{f}). The inset in \textbf{d} is a zoom on the region delimited by the black dotted square.
\textbf{g, h, i,} Measured $f(E)$ for the configurations in the 480~nm device depicted in resp. \textbf{a}, \textbf{b}, and \textbf{c}. Each curve, offset for clarity, corresponds to an injection energy increment $\delta E_1\approx 21~\mu$eV.
}
\end{figure*}

Our analysis shows that the two devices differ not only by their nominal length and external dissipation, but also by their plasmon velocities. Indeed, the $750~$nm device presents a larger difference between $v_\rho$ and $v_\sigma$, effectively increasing the energy relaxation (or, in other words, making it effectively much longer than the 480~nm device as far as energy relaxation due to EC coupling is concerned.) To interpolate between those two different cases, we rely on the length gate, the basic effect of which is illustrated in Fig.~\ref{fig6}. For positive $\Vlength\approx 0.2~$V, the gate does not affect the trajectory of the edge channels, which flow straight from QD1 to QD2 (Fig.~\ref{fig6}a). The corresponding $f(E)$ measured for $L=750~$nm are shown in Fig.~\ref{fig6}d, and are similar to the data shown in Fig.~\ref{fig2}. For intermediate values $\Vlength\approx -0.1~$V, the electrostatic potential generated by the gate allows separating the two ECs \cite{Inoue2014}, as depicted in Fig.~\ref{fig6}b: spectacularly, in that case all data show a very clear quasiparticle peak up to large $E_1$ (Fig.~\ref{fig6}e). In contrast, for large negative values $\Vlength\approx -0.5~$V, both edge channels are diverted around the gate and follow a longer path ($L=2.17~\mu$m, Fig.~\ref{fig6}c), leading to the full disappearance of the quasiparticle peak even at low $E_1$ (Fig.~\ref{fig6}f, see also Fig.~\ref{fig5}). The quasiparticle peak evolution in the data shown in Fig.~\ref{fig6}d and \ref{fig6}e can be reproduced using our model (see Supplementary Note 5), with slightly different Fermi velocities for the two datasets (but the same velocity ratio $\alpha=2.1$). Interestingly, while the friction coefficient $\gamma_0=0.13$ is the same for the two datasets (as well as for the other measurements in the $L=750~$nm device), the extracted EC coupling $u\approx\{83, 21\}~\mathrm{km}~\mathrm{s}^{-1}$ is four times smaller when the two channels are separated. The length gate on the 480 nm device makes it possible to manipulate the ECs in the same way (Fig.~\ref{fig6}g-i), allowing us to separate the ECs Fig.~\ref{fig6}h), as well as to increase the co-propagation length to $L\approx1.3~\mu$m (Fig.~\ref{fig6}i). For the latter length, the quasiparticle peak decreases sharply, but remains visible up to $100~\mu$eV. The TLL analysis of both datasets shows that, as for the 750~nm device, the friction coefficient remains constant, $\gamma_0=0.43$ (see Supplementary Note 5 for additional plots and TLL analysis). 
For smaller gate voltages that do not fully separate the ECs, the length gate, coupled to the central gate separating the two QDs (see Fig.~\ref{fig1}a), can nevertheless modify the electrostatic potential that defines the ECs flowing between the two dots, thereby granting us an additional control over the TLL parameters. We have performed the spectroscopy and TLL analysis of the quasiparticle peak height on the 480~nm and 750~nm devices for various gating configurations (see Supplementary Note 5 for plots and analysis, as well as a table summarizing the extracted TLL parameters). We observe consistently that the gate configuration allows tuning the plasmon velocities $v_\rho$ and $v_\sigma$, while the friction coefficient $\gamma_0$ remains constant in each device.

\subsection{Prethermalization}

\begin{figure}[h]
\centering
\includegraphics[width=0.45\textwidth]{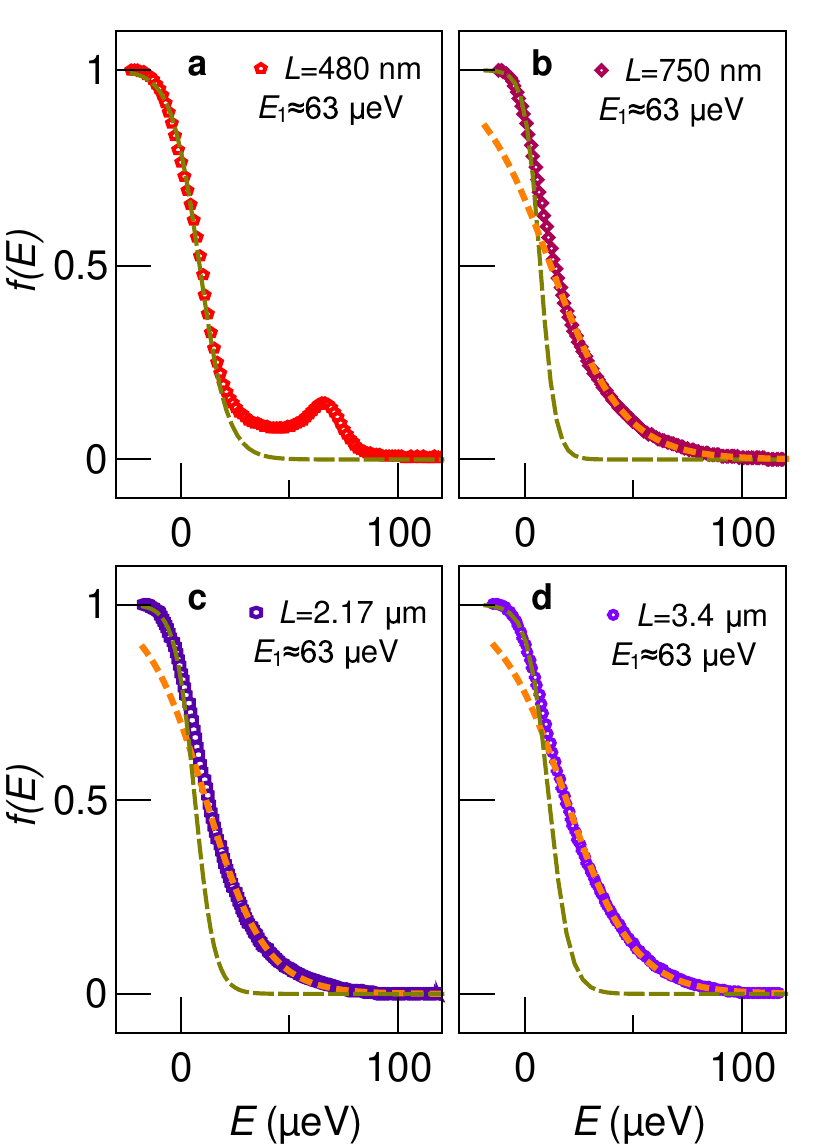}
\caption{\label{fig5} \textbf{$\vert$ Length dependence and prethermalization.} Measured $f(E)$ for an injection energy $E_1\approx63~\mu$eV at \textbf{a)} $L=480~$nm, \textbf{b)} $L=750~$nm, \textbf{c)} $L=2.17~\mu$m, and \textbf{d)} $L=3.4~\mu$m. The symbols are experimental data, and the dotted (resp. dashed) lines are Fermi function fit of the high energy (resp. low energy) part of the data.
}
\end{figure}

We finally turn to the evolution of the measured distribution functions for propagation lengths above $1~\mu$m. The integrability of the TLL model (in absence of external dissipation) implies that energy relaxation should not lead to an equilibrium Fermionic state described by a high temperature Fermi function \cite{Takei2010, Levkivskyi2012}. This property has been recently confirmed by the observation of \textit{prethermalized} states after the relaxation of highly imbalanced double step distribution functions created by a biased QPC \cite{Itoh2018}, but, up to now, not for finite energy quasiparticles. We have observed that as the propagation length is further increased, the quasiparticle peak fully vanishes. Notably, when the peak is no longer visible, the distribution function does not qualitatively change, up to our longest studied length, $L=3.4~\mu$m. This is illustrated in Fig.~\ref{fig5} (see also Fig.~\ref{fig6}f), where we have plotted the measured $f(E)$ corresponding to the same injection energy $E_1\approx63~\mu$eV. Apart from the data at $L=480~$nm, which display a clear quasiparticle peak, all other lengths yield similar, monotonous $f(E)$. These distribution functions cannot be fully fitted by a Fermi function: the dashed and dotted lines in Fig.~\ref{fig5} are tentative fits of the (respectively) low- and high-energy part of the distribution functions, with significantly different effective temperature for the high energy part ($\sim160-195~$mK) with respect to the low energy part ($\sim40-55~$mK - see also Supplementary Note 6). Both fits show significant deviations with respect to the data. As a sanity check, we have measured equilibrium distribution functions at elevated temperatures ($T\approx160~$mK), corresponding to energy width similar to the data shown in Fig.~\ref{fig5}, which showed much smaller deviations to a Fermi function (see Supplementary Note 6). Despite the significant role of energy losses towards external degrees of freedom, which should lead to a thermalized state after long propagation length, this apparent long-lived nonthermal behavior could indeed be a signature of TLL prethermalization. Furthermore, this might explain the recently reported robust quantum coherence of finite energy quasiparticles emitted in a Mach-Zehnder interferometer \cite{Tewari2016}. The apparent competition between prethermalization and observed dissipation is highly intriguing, and beckons further theoretical investigation of the impact of dissipation in the TLL model.

\section{Discussion}

To summarize, we have directly observed the relaxation and revival of quasiparticles emitted at finite energy in an edge channel at filling factor $\nu=2$ of the quantum Hall effect. These results qualitatively reproduces the hallmark phenomenology of the TLL model, and we show that the quantitative discrepancies are well accounted for by introducing dissipation in the model. In order to maximize the phase coherence and energy relaxation lengths in electron quantum optics experiments, one should not only rely on schemes that limit the effect of inter EC coupling \cite{Altimiras2010, Huynh2012, Cabart2018, Duprez2019}, but also identify the mechanisms behind this dissipation. A possible cause of this dissipation could be the recently observed long distance Auger-like processes \cite{Krahenmann2019}, although their signature is again not visible in our data. This stresses the need for further research in order to fully grasp the physics of interactions at $\nu=2$ \cite{Goremykina2019}.

\section{Acknowledgements}
We warmly thank C. Altimiras, A. Anthore, G. F\`eve, F. Pierre, E. Bocquillon, X. Waintal and P. Degiovanni for enlightening discussions, as well as P. Jacques for technical support. This work was funded by the CEA and the French RENATECH program.

\section{Author contributions}
R.H.R. performed the experiments, with help from F.D.P., P.R. \& Pat.R.; R.H.R., F.D.P., D.F., \& Pat.R. analyzed the data, with inputs from UG., D.M. \& F.P.; D.F. \& M.S. developed the theoretical model; U.G. \& A.C. grew the 2DEG; D.M. fabricated the devices, with inputs from R.H.R., F.D.P. \& Pat.R.; R.H.R., F.D.P., U.G., D.M. \& Pat.R. wrote the manuscript, with inputs from all other coauthors; Pat.R. supervised the project.

\section{Competing Interests}
The authors declare no competing interests.

\section{Methods}
\subsection{Samples}
The samples were realized in a 90~nm deep GaAs/GaAlAs two-dimension electron gas (2DEG), with typical density $\sim 2.5\times10^{11}~\mathrm{cm}^{-2}$ and mobility $\sim 2\times10^6~\mathrm{cm}^2~\mathrm{V}^{-1}~\mathrm{s}^{-1}$, cooled down to electronic temperatures of $\sim 20-30~$mK. Perpendicular magnetic fields of about 5~T were applied to reach filling factor $\nu=2$ of the quantum Hall effect. 
\subsection{Measurements}
Measurements were performed in a dilution refrigerator, using standard low frequency lock-in techniques. For each configuration of the experiment, the drain and source voltages $\VD$ and $\VS$ are tuned such that only a single narrow resonance sits in the transport window, with no excited states present. The spectroscopy range is then set by the minimum of $\{\mathrm{e}\VD,\vert \mathrm{e}\VS\vert\}$.
\subsection{Additional checks}
To ensure that no tunneling takes place between the two copropagating edge channels, we check that the elevation of the electrochemical potential in the outer edge channel, obtained by integrating the measured $f(E)$, is equal to its expected value (see Supplementary Note 4).

\section{Data availability}
The data and analysis used in this work are available from the corresponding author upon reasonable request.

\cleardoublepage



\section*{Supplementary material}

# Supplementary Information for
# Relaxation and revival of quasiparticles injected in an interacting quantum Hall liquid

R. H. Rodriguez,[1] F. D. Parmentier,[1] D. Ferraro,[2,3] P. Roulleau,[1] U. Gennser,[4] A. Cavanna,[4] M. Sassetti,[2,3] F. Portier,[1] D. Mailly,[4] and P. Roche[1]

[1]*Université Paris-Saclay, CEA, CNRS, SPEC, 91191, Gif-sur-Yvette, France*
[2]*Dipartimento di Fisica, Università di Genova, Via Dodecaneso 33, 16146, Genova, Italy*
[3]*SPIN-CNR, Via Dodecaneso 33, 16146, Genova, Italy*
[4]*Centre de Nanosciences et de Nanotechnologies (C2N), CNRS, Université Paris-Sud, Université Paris-Saclay, 91120 Palaiseau, France*

(Dated: April 17, 2020)

## SUPPLEMENTARY NOTE 1: MEASURED SAMPLES

### Devices and datasets

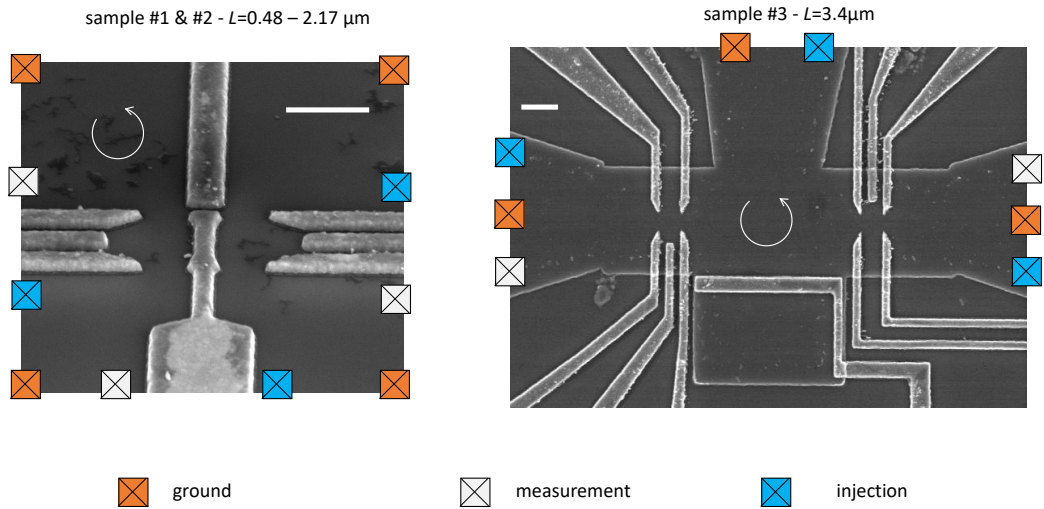

**Supplementary Figure 1 | Samples.** Scanning electron micrographs of typical sample geometries (left: samples #1 and #2, $L = 0.48 - 2.17$ µm; right: sample #3, $L = 3.4$ µm). The circular arrow indicate the chirality of the edge channels. The white scale bars correspond to 500 nm. The ohmic contacts are symbolized by the crossed squares, the color of which indicate the contact's role in the measurement circuit (orange: cold ground, grey: ac and dc current feed, blue: measurement).

All data discussed in the main text were measured in three samples (#1, #2 and #3). Typical geometries, as well as ohmic contact layout and configuration, are shown in Supplementary Fig. 1. A complete set of curves (referred to in the following as a spectrum), as shown in the figures of the main text is obtained using a unique pair of resonances of the QDs. Supplementary table 1 summarizes the resonances implemented in the emitter (QD1) and the detector (QD2) for each spectrum (S1 to S8), the corresponding figure of the main text, the propagation length $L$, the respective transmission $T_{1,2}$ and width $\Gamma_{1,2}$ of emitter and detector resonances (see below), the characteristic decay energy $E_d$, and the average width (FWHM) of the detected QP peak.

| Sample | Data | Res-QD1 | Res-QD2 | Fig. | L | $T_1$ | $\Gamma_1$ | $T_2$ | $\Gamma_2$ | FWHM |
|---|---|---|---|---|---|---|---|---|---|---|
| | | | | | $\mu$m | | $\mu$eV | | $\mu$eV | $\mu$eV |
| #1 | S1 - 480 nm res. A | 1A | 2A | 2, 3a, 4, 6 | 0.48 | 0.44 | 12.8 | 0.3 | 5.9 | 19 |
| #1 | S2 - 480 nm res. B | 1B | 2B | 4, 6 | 0.48 | 0.60 | 18.7 | 0.50 | 11.25 | 30 |
| #2 | S3 - 750 nm cooldwn 1 | 1C | 2C | 2, 3b, 4 | 0.75 | 0.54 | 16.3 | 0.34 | 9.6 | 29 |
| #2 | S4 - 750 nm cooldwn 2 | 1C | 2C | 2, 3b, 4 | 0.75 | 0.50 | 15.0 | 0.50 | 10.0 | 22 |
| #2 | S5 - 750 nm short | 1C | 2C | 5d | 0.75 | 0.9 | 15.0 | 0.19 | 9.8 | 29 |
| #2 | S6 - 750 nm sep. | 1D | 2D | 5e | 0.75* | 0.40 | 11.8 | 0.55 | 4.9 | 18 |
| #2 | S7 - 750 nm long | 1E | 2E | 5f, 6 | 2.17 | 0.43 | 17.0 | 0.20 | 27.0 | – |
| #3 | S8 - 3400 nm | 1F | 2F | 6 | 3.4 | 0.25 | 19.75 | 0.43 | 8.0 | – |

**Supplementary Table 1** | Measured samples and the corresponding data discussed in the main text. The column Res-QD1 (Res-QD2) refers to the resonance in the emitter (detector) QD used to measure each spectrum. Spectrum S6 corresponds to the $L = 750$ nm dataset with separated ECs.

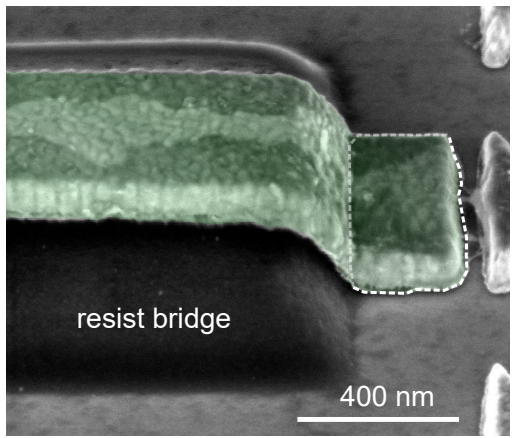

**Supplementary Figure 2** | **Length gate.** Scanning electron microscope side view of the side gate, showing the insulating SU-8 resist bridge separating the gate from the surface of the sample. The imprint of the gate on the 2DEG is highlighted by the white dashed line.

### Length gate

A side view of the length gate is shown in Supplementary Fig. 2, displaying the 200 nm-thick SU-8 resist layer separating the left part of the length gate from the surface of the sample, greatly weakening it electrostatic influence on the 2DEG. As a result, when applying a voltage to that gate, only the $\approx 300x500$ nm region of the 2DEG at the end of the gate (circled with a white dashed line in Supplementary Fig. 2) sees a change in carrier density. Without this resist bridge, the two quantum dots would become electrically isolated by the depleted region extending all across the 2DEG.

Supplementary Fig. 3 depicts how the short and long path lengths are extracted from scanning electron micrographs for the 750 nm device. The same method is used to extract the lengths for the 480 nm device.

### SUPPLEMENTARY NOTE 2: QUANTUM DOTS CALIBRATION

The resonances used in the emitter ($i = 1$) and detector ($i = 2$) QDs were characterized by measuring the transconductance $\partial I_i/\partial V_{\text{Pi}}$ as a function of the respective bias voltage ($V_D$ for $i = 1$ and $V_S$ for $i = 2$) and the plunger gate voltage $V_{\text{Pi}}$. It is in particular crucial to work with resonances devoid of excited states, so as to be sure that quasi-particles are emitted at a single well defined energy [1]. These usually are easily identified when measuring Coulomb diamonds plots. The color plot in Supplementary Fig. 4 shows a typical measurement in the sequential tunneling

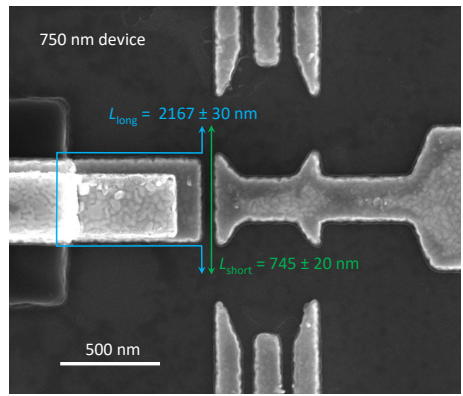

**Supplementary Figure 3 | Estimation of the propagation lengths on the 750 nm sample.** Scanning electron microscope view of the sample, with the green arrow depicting the short path ($L = 750$ nm) and the blue arrow depicting the long path ($L = 2.17$ $\mu$m) circumventing the length gate.

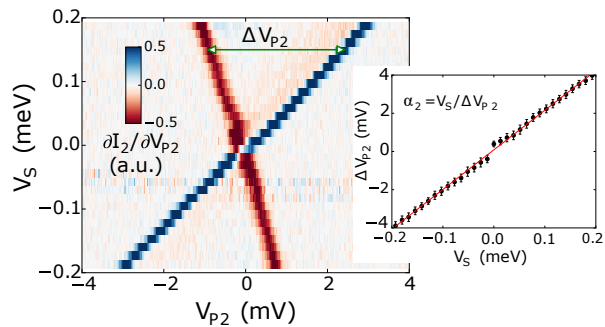

**Supplementary Figure 4 | QD calibration.** The color map is a typical transconductance measurement of a QD in the Coulomb blockade regime. The inset is the linearisation from which we obtain the lever arm $\alpha$.

regime. The red and blue oblique lines, separated by $\Delta V_{\text{Pi}}$, define the boundaries of the Coulomb blockade regime. A linear fit of $\Delta V_{\text{Pi}}$ as a function of the bias voltage allows us to extract the lever arm $\alpha_i$, as shown in the plot at the right hand side. Supplementary table 2 summarizes the lever arm obtained from the calibration of each of the resonances used in this work. The transmission and widths of the resonances (see Supplementary Table 1) are extracted from the differential conductance $\partial I_i/\partial V_{\text{D,S}}$ at zero bias voltage. In our experiment, we have tried to work, for both QDs, with very narrow resonances (with widths ideally smaller than the temperature) so as to maximize the energy resolution, and with close to unity transmission (corresponding to symmetric barriers), so as to maximize the signal-to-noise ratio. In practice, this turns out to be quite challenging, especially when dealing with quantum dots sharing a common depletion gate. The data shown in the main text and in the supplementary information correspond to the best experimental conditions we could achieve. When measuring the distribution function $f(E)$ obtained by sweeping $V_{\text{P2}}$, we have simultaneously corrected for the cross-talk between $V_{\text{P1}}$ and $V_{\text{P2}}$, in order to keep a constant injection energy $E_1(V_{\text{P1}})$. Typically, for a step $\delta V_{\text{P2}}$ in QD2, we correct the plunger gate voltage in the QD1 by $\delta V_{\text{P1}} \approx -\delta V_{\text{P2}}/20$.

## SUPPLEMENTARY NOTE 3: CONVOLUTION

The transmitted current $I_2(E_2)$ trough the detector QD2 in the sequential tunneling regime reads:

$$I_2(E_2) = \frac{e}{h} \int L_2(E, E_2) \big[f(E) - f_S(E)\big] dE \tag{1}$$

where $f(E)$ is the distribution function to be probe, $f_S(E)$ the equilibrium distribution function at the source of

| Spectrum | $\alpha_1$ $(10^{-2}$ meV/mV$)$ | $\alpha_2$ $(10^{-2}$ meV/mV$)$ |
|---|---|---|
| S1 | $3.64 \pm 0.02$ | $3.97 \pm 0.02$ |
| S2 | $3.55 \pm 0.03$ | $0.406 \pm 0.001$ |
| S3 | $4.80 \pm 0.03$ | $6.35 \pm 0.03$ |
| S4 | $4.80 \pm 0.03$ | $6.35 \pm 0.03$ |
| S5 | $4.80 \pm 0.03$ | $6.35 \pm 0.03$ |
| S6 | $4.79 \pm 0.03$ | $4.97 \pm 0.02$ |
| S7 | $4.09 \pm 0.02$ | $3.97 \pm 0.02$ |
| S8 | $1.14 \pm 0.02$ | $3.28 \pm 0.02$ |

**Supplementary Table 2** | Measured lever arms $\alpha_i$ for the resonances used in each QD ($i = 1, 2$) to obtain the spectra discussed in the main text.

the detector, and $L_2(E, E_2)$ is the intrinsic lineshape of the resonance on QD2. A simple model of electrostatic confinement for the electrons in the QD gives $L_2(E, E_2)$ as a Lorentz function centered in $E_2$.

If the resonance of the detector is narrow enough, namely its intrinsic width $\Gamma_2 \ll k_B T$, the lineshape can be approximated by a normalized delta function $L_2(E, E_2) \sim L_0 \times \delta(E - E_2)$, with $L_0 = \int L_2(E, E_2) dE$ a constant characteristic of the detector. Thus the current $I_2(E_2)$ is directly proportional to $\Delta f(E) = f(E) - f_S(E)$.

In other cases, as for most of the resonance used in our experiment, the measured signal, obtained from $\partial I_2/\partial V_{P2}$, is convoluted with the lineshape of QD2 following Eq. 1. As a consequence, the detected Fermi sea is widened, and the effective electronic temperature $T_{eff}$, that is obtained by fitting a Fermi function, is larger than the 2DEG electronic temperature $T$, which is measured with a much narrower resonance. This is the case of the Fermi sea in all the spectra presented in the main text. The table 3 summarizes the measured $T_{eff}$, $T$ and the base temperature $T_{ph}$ for the different spectra. A comparison between $T_{eff}$ and $T$ provides a method to estimate the linewidth $\Gamma_2$ of the detector. An equivalent approach can be followed to further characterize the emitter QD.

| Spectrum | $T_{eff}$ (mK) | $T$ (mK) | $T_{ph}$ (mK) |
|---|---|---|---|
| S1 | $40 \pm 2$ | $23 \pm 1$ | $18.5 \pm 0.2$ |
| S2 | $51 \pm 2$ | $23 \pm 1$ | $18.1 \pm 0.1$ |
| S3 | $48 \pm 2$ | $30 \pm 1$ | $17.9 \pm 0.1$ |
| S4 | $67 \pm 4$ | $30 \pm 1$ | $18.0 \pm 0.1$ |
| S5 | $48 \pm 2$ | $30 \pm 1$ | $17.9 \pm 0.1$ |
| S6 | $39 \pm 2$ | $27 \pm 2$ | $15.9 \pm 0.4$ |
| S7 | $59 \pm 3$ | $19 \pm 1$ | $17.5 \pm 0.1$ |
| S8 | $51 \pm 2$ | $30 \pm 5$ | $15.3 \pm 0.1$ |

**Supplementary Table 3** | Comparison between the effective electronic temperature $T_{eff}$, the 2DEG electronic temperature $T$ and the bath temperature $T_{ph}$.

The convoluted distribution shows also some deviation from an actual Fermi function, mainly by the development of a long tail. In such a case the convoluted distribution function is better described by an *arctangent* function with a characteristic width $T_{atn}$:

$$f(E) \approx \frac{1}{2} - \frac{1}{\pi} Arctan\left(\frac{E}{k_B T_{atn}}\right) \qquad (2)$$

Notice that this functional form is the same as that predicted for a metastable state, which is expected to occur in the relaxation process of a double step distribution function generated by a QPC at low transmission [2]. Thus the effects of the convolution can hamper the experimental investigation of the predicted metastable state [3].

Moreover, the injected QP peak in our experiment is also affected by the convolution at the detector QD. Let us consider the case when the injected QP peak is a Lorentz peak $L_1(E, E_1)$ centered at the injection energy $E_1$, with amplitude $\mathcal{T}_1$ and width $\Gamma_1$. After the convolution with the QD2 lineshape, $L_2(E, E_2)$, the QP peak maintains its

Lorentzian form but with an increased width: $\Gamma_1 + \Gamma_2$, and a reduced height: $\mathcal{T}_1\Gamma_1/(\Gamma_1 + \Gamma_2)$. Importantly, since the characteristics of the detector are the same for all the curves of the same spectra, the exponential decay of the peak height discussed in the main text is not affected by the convolution. As can be seen in Supplementary Table I, the extracted widths of the Lorentzian fits correspond within less than 10 % deviations to the sum $\Gamma_1 + \Gamma_2$.

## SUPPLEMENTARY NOTE 4: ELECTROCHEMICAL POTENTIAL SHIFT

An important check in our experiments consists in verifying that the charge current remains conserved in the edge channel, *i.e.* that no charge tunnels from one edge channel to the other. This can be done by calculating the electrochemical potential shift $\Delta\mu$ in the outer edge channel at the detector, given by the integral of the measured distribution function, and comparing it either to the amount of current $I_1$ stemming from the drain that is transmitted by QD1 (note that in practice we measure the reflected current $1 - I_1$), or to the shift given by the emitter's resonance parameters, namely $\Delta\mu = \frac{\pi}{2}\Gamma_1 T_1$ for a Lorentzian shaped resonance. This is illustrated in Supplementary Fig. 5 for the $L = 480$ nm data shown in main text Fig. 2. We systematically observe a good agreement, indicating that charge current is always conserved in the outer edge channel in our experiments.

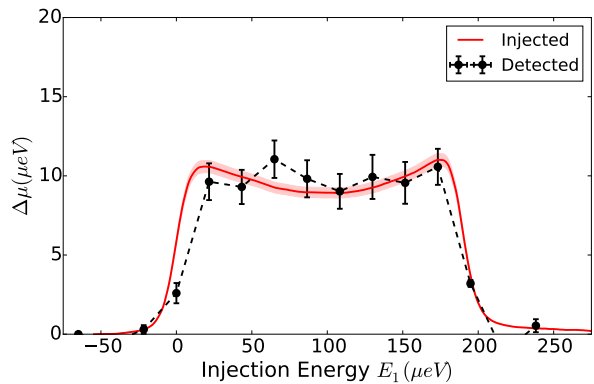

**Supplementary Figure 5 | Electrochemical potential shift for $L = 480$ nm.** Symbols: electrochemical potential shift $\Delta\mu$ after propagation, obtained by integrating the measured $f(E)$ shown in main text Fig. 2, versus $E_1$. The error bars correspond to the combined uncertainties on the QDs lever arms and on the normalization of the $f(E)$. Red line: $\Delta\mu$ at the injection, determined from the current flowing through QD1. The red shaded region corresponds to uncertainties on the QDs lever arms. The value of $\Delta\mu$ extracted from the resonance of QD1 is $\frac{\pi}{2}\Gamma_1 T_1 \approx 8.8$ $\mu$eV.

## SUPPLEMENTARY NOTE 5: ADDITIONAL DATA AND ANALYSIS

We present here the measured distribution functions and their TLL analysis for additional spectra obtained on the 480 nm and 750 nm devices. The spectra (as well as the one shown in the main text) are labelled R**xx**, where the number **xx** indicates the order in which the data was obtained over the whole measurement run. Spectra R16-R21 where obtained during a first cooldown of the 750 nm device, and spectra R24-R48 during a subsequent cooldown of the same device. Spectra R72-R106 were obtained during a single cooldown of the 480 nm device. The missing numbers correspond to incomplete datasets (*e.g.* due to gate instabilities, or malfunctions of the dilution refrigerator).

Supplementary Fig. 6 shows the spectrum S2, obtained for a slightly broader resonance (labelled resonance B) of the $L = 480$ nm sample, in semi-log scale, along with the Lorentzian fits, the normalized heights of which are shown as black hexagons in main text Fig. 4 (we recall that in the main text, the peak height is normalized by height of the injected quasiparticle peak, given by the transmission of the first quantum dot). The peak positions and FWHM are shown in the inset. Note that in our analysis, we do not remove any background in the distribution functions (regardless of the dataset) before performing the Lorentzian fits. Supplementary Fig. 7 shows the spectrum S4, obtained for a second cooldown (labelled cooldown 2) of the $L = 750$ nm sample, corresponding to the data shown as red and orange circles in main text Fig. 4. We also show in Supplementary Fig. 8 the data and fits appearing in main text Fig. 4, in linear scale. The discrepancy between the data and the fits at small $E_1 \sim 20$ $\mu$eV can be accounted for

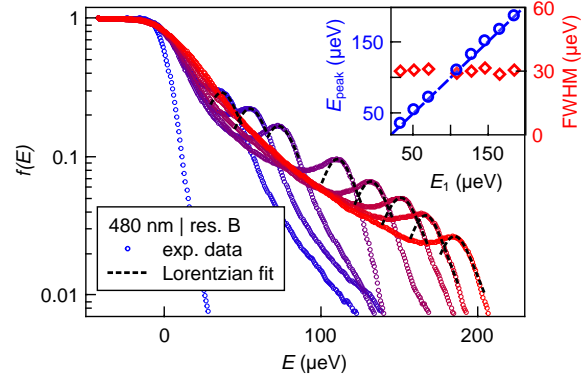

**Supplementary Figure 6 | Quasiparticle peak analysis for spectrum S2:** $L = 480$ **nm, resonance B.** Measured $f(E)$ in sample #1, plotted in semi-log scale. The inset shows the peak center (blue circles) and the peak width (red diamonds), determined from the Lorentz fit (black dash line) of the peak, as a function of the injection energy $E_1$.

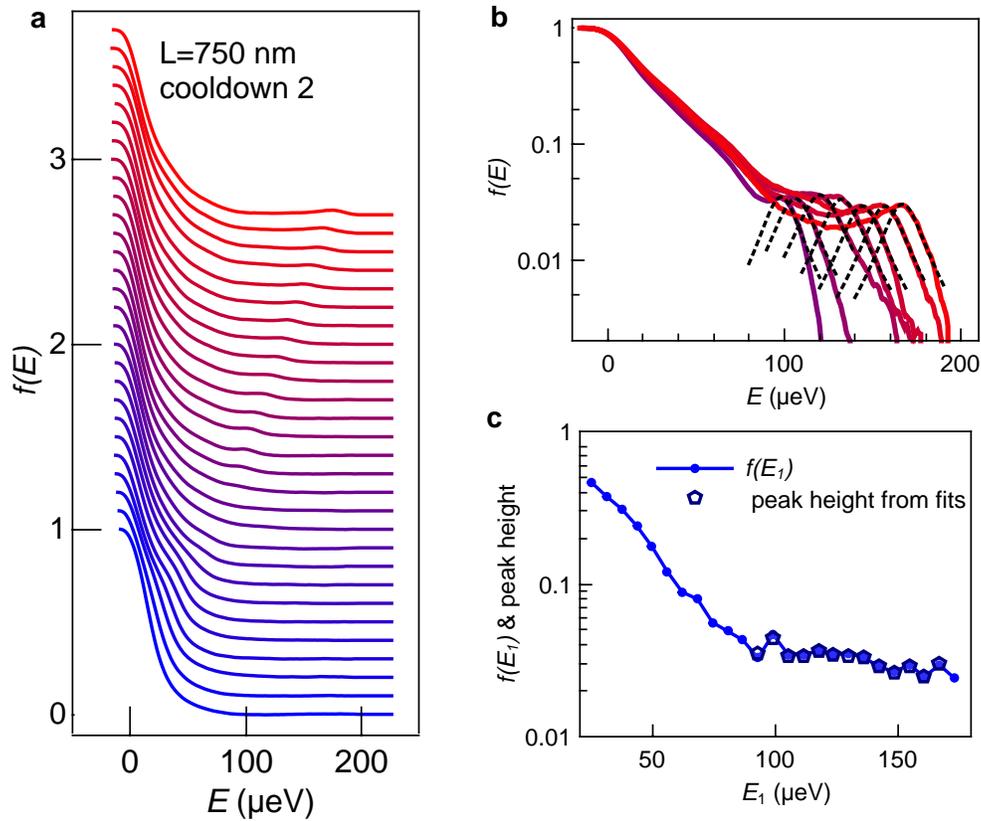

**Supplementary Figure 7 | Quasiparticle peak analysis for spectrum S4:** $L = 750$ **nm, cooldown 2. a,** Measured $f(E)$. Each curve, offset for clarity, corresponds to an increment of the injection energy $\delta E_1 \approx 6\mu$eV, from $E_1 \approx 6\mu$eV (blue), to $E_1 \approx 173\mu$eV (red). **b,** Measured $f(E)$, plotted in semi-log scale. The black dashed lines are Lorentzian fits of the quasiparticle peak. **c,** Peak height versus $E_1$, in semi-log scale. The full blue circles are extracted $f(E_1)$, and the pentagons the peak heights extracted from the Lorentzian fits.

by the fact that at low energy, the quasiparticle peak sits on the tail of the Fermi distribution function corresponding to the Fermi sea of the EC.

Supplementary Fig. 9 shows the spectrum S5, obtained at $L = 750$ nm when the ECs are not separated (see main

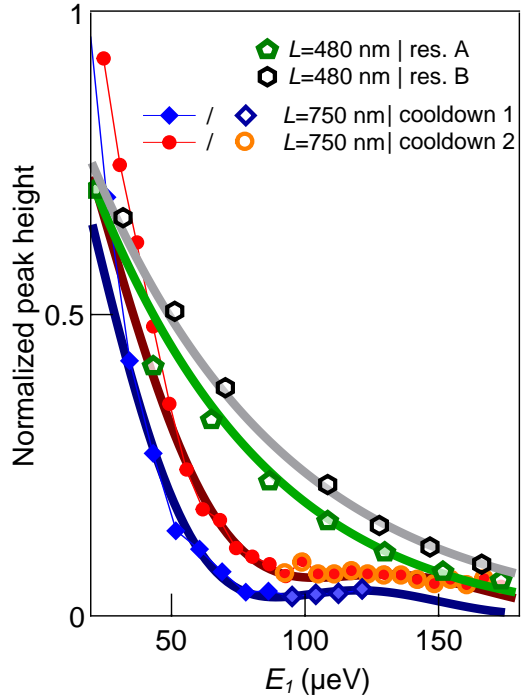

**Supplementary Figure 8 | Normalized peak height.** Normalized peak height versus $E_1$ (same data as in main text Fig. 4), in a linear scale. The fitting parameters are reported in Supplementary Table IV.

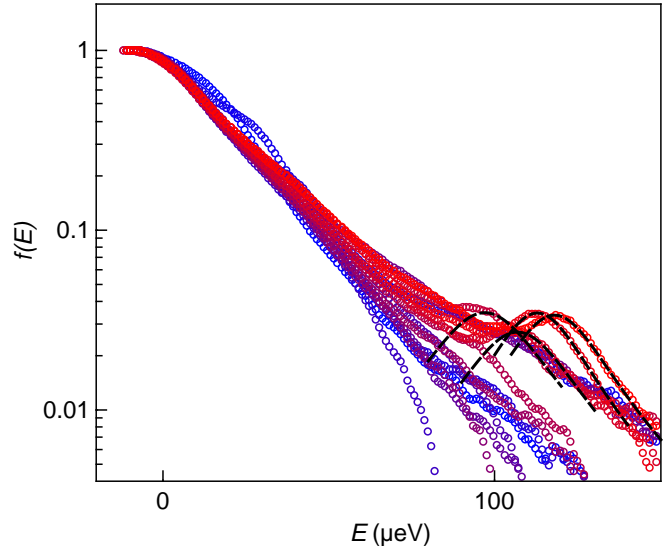

**Supplementary Figure 9 | Quasiparticle peak analysis for spectrum S5:** $L = 750$ **nm with copropagating ECs.** Measured $f(E)$ in semi-log scale (symbols), with Lorentzian fits (black dashed lines).

text Fig. 5d), in semi-log scale, among with Lorentzian fits of the high $E_1$ data where the quasiparticle peak is visible again. The data at $L = 750$ nm when the ECs are separated (see main text Fig. 5e) is shown in semi-log scale in Supplementary Fig. 10. Along with the peak position and FWHM, we show the evolution of the peak height versus $E_1$. Note that in this case the exponential character of the decay is less clear, as the peak height only decreases by 50 %. We show in Supplementary Fig. 11 the fits of quasiparticle peak height obtained from the above data (750 nm, copropagating and separated ECs), using our refined TLL model. The TLL parameters are summarized in Supplementary Table IV.

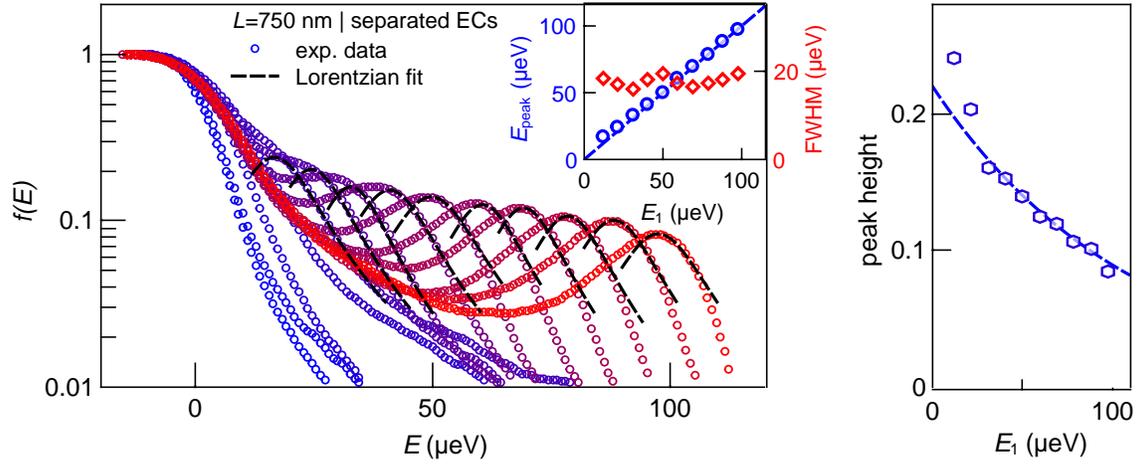

**Supplementary Figure 10 | Quasiparticle peak analysis for spectrum S6: $L = 750$ nm with separated ECs.** Left panel: measured $f(E)$ in semi-log scale (symbols), with Lorentzian fits (black dashed lines). The extracted peak positions and FWHM are shown in the inset. Right panel: absolute peak heights (blue hexagons) extracted from the Lorentzian fits shown in the left panel, plotted in linear scale versus $E_1$. The blue dashed line is the an exponential fit. The fitting parameters are reported in Supplementary Table IV.

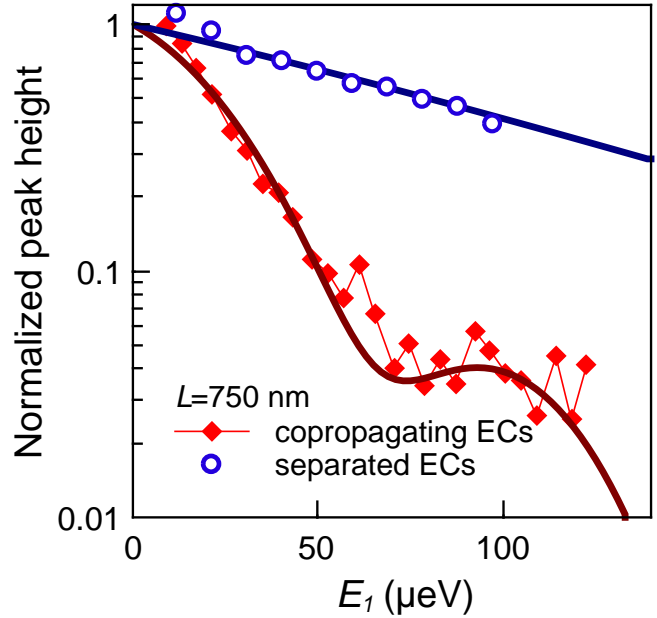

**Supplementary Figure 11 | TLL fits of the 750 nm data with copropagating (spectrum S5) /separated (spectrum S6) ECs.** Normalized peak height versus $E_1$. Red diamonds: extracted $f(E_1)$ for copropagating ECs. Blue circles: peak heights from Lorentz fits for separated ECs). The thick lines are fits using the TLL model with dissipation. The fitting parameters are reported in Supplementary Table IV.

Supplementary Figures 12, 13, 14, 15, 16 and 17 show the measured distribution functions as well as the corresponding analysis for the datasets listed in Fig. 6 of the main paper (for both 480 and 750 nm devices), in particular the long path (Fig. 12) and the separated ECs (Fig. 13) configurations of the 480 nm device. The parameters extracted from the TLL analysis are summarized in Supplementary Table IV.

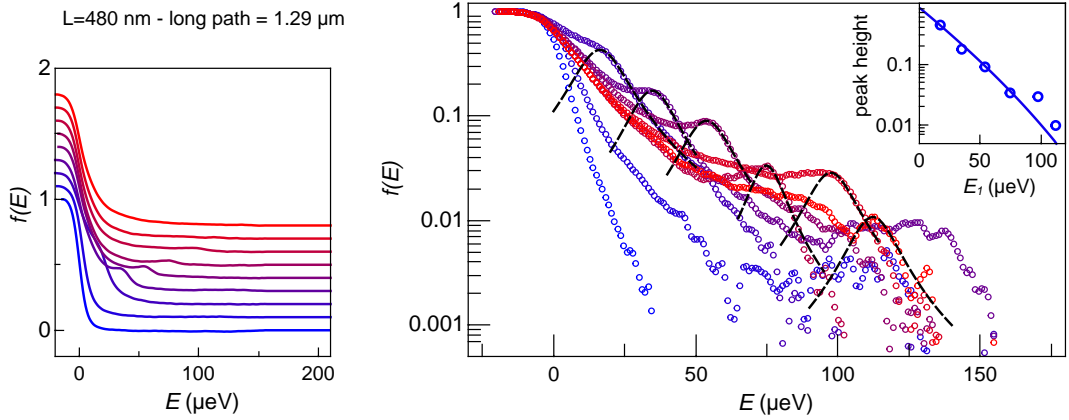

**Supplementary Figure 12 | TLL analysis for additional spectrum R76: long path of the 480 nm device.** Left Panel: measured $f(E)$. Each curve, offset for clarity, corresponds to an increment $\delta E \approx 19$ $\mu$eV of the injection energy. Right panel: measured $f(E)$ in semi-log scale (symbols), with Lorentzian fits (black dashed lines). Inset: peak heights (blue circles) extracted from the Lorentzian fits shown in the left panel, plotted in semi-log scale versus $E_1$. The blue line is a fit using the TLL model with dissipation. The fitting parameters are reported in Supplementary Table IV.

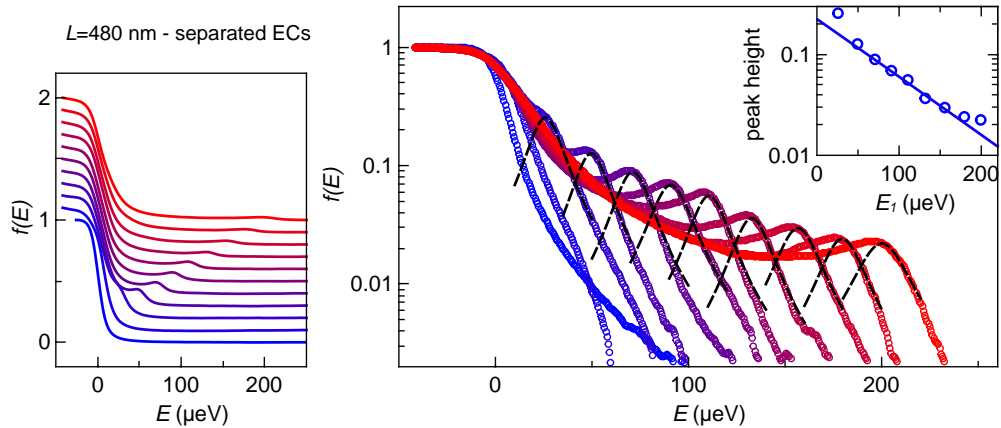

**Supplementary Figure 13 | TLL analysis for additional spectrum R84: 480 nm device with separated ECs.** Left Panel: measured $f(E)$. Each curve, offset for clarity, corresponds to an increment $\delta E \approx 22$ $\mu$eV of the injection energy. Right panel: measured $f(E)$ in semi-log scale (symbols), with Lorentzian fits (black dashed lines). Inset: peak heights (blue circles) extracted from the Lorentzian fits shown in the left panel, plotted in semi-log scale versus $E_1$. The blue line is a fit using the TLL model with dissipation. The fitting parameters are reported in Supplementary Table IV.

## SUPPLEMENTARY NOTE 6: LENGTH DEPENDENCE AND THERMALIZED STATE

At lengths larger than 750 nm, the quasiparticle peak fully vanishes and the measured distribution functions become monotonous. Main text Fig. 6 shows that these present systematic discrepancies with respect to Fermi functions. In particular, fitting the high energy part of the distribution function with a Fermi function systematically leads to an excess of particles at low energy. Measuring equilibrium $f(E)$ (that is, when no additional quasiparticle is emitted by the first quantum dot) at high temperature ($T = 157$ mK, see Fig. 19a) yields much smaller low-energy deviations to a Fermi function (note that in that case the temperature is not a fitting parameter, and is fixed to the actual electron temperature extracted from the characterization of the quantum dots). Note that there is no theoretical prediction for the shape of prethermalized state obtained after the relaxation of finite-energy quasiparticle in addition to a finite temperature Fermi function (such a prediction only exists so far for a biased quantum point contact with extreme

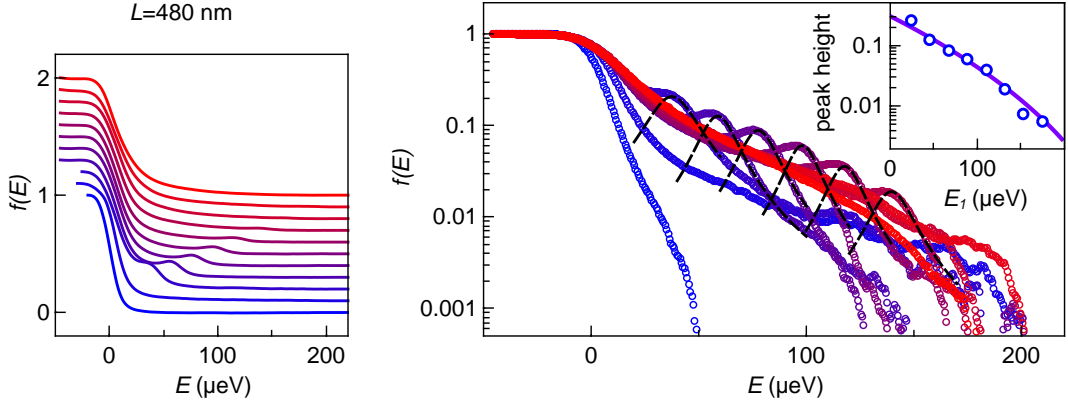

**Supplementary Figure 14 | TLL analysis for additional spectrum R99: 480 nm device** Left Panel: measured $f(E)$. Each curve, offset for clarity, corresponds to an increment $\delta E \approx 21$ µeV of the injection energy. Right panel: measured $f(E)$ in semi-log scale (symbols), with Lorentzian fits (black dashed lines). Inset: peak heights (blue circles) extracted from the Lorentzian fits shown in the left panel, plotted in semi-log scale versus $E_1$. The blue line is a fit using the TLL model with dissipation. The fitting parameters are reported in Supplementary Table IV.

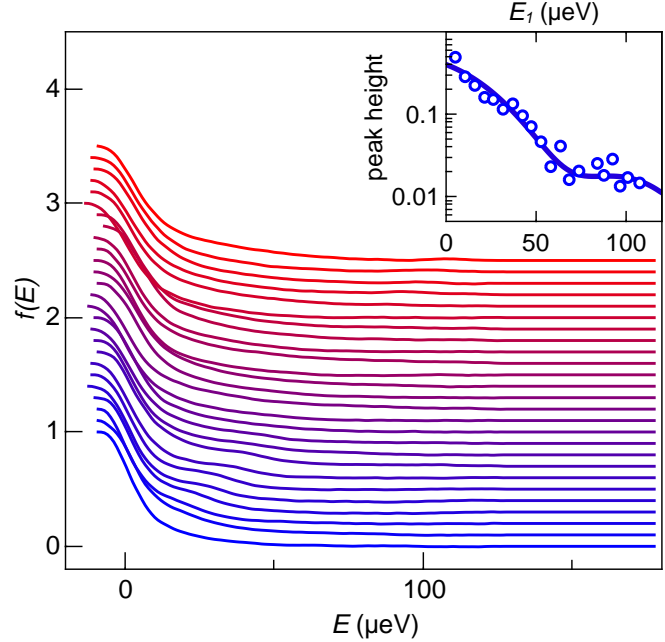

**Supplementary Figure 15 | TLL analysis for additional spectrum R16: 750 nm device.** Measured $f(E)$. Each curve, offset for clarity, corresponds to an increment $\delta E \approx 5$ µeV of the injection energy. Inset: peak heights $f(E_1)$ (blue circles) plotted in semi-log scale versus $E_1$. The blue line is a fit using the TLL model with dissipation. The fitting parameters are reported in Supplementary Table IV.

transmission [2, 3]).

### SUPPLEMENTARY NOTE 7: COMPARISON WITH PREVIOUS EXPERIMENTS USING DOUBLE STEP DISTRIBUTION FUNCTIONS

It is not straightforward to make direct quantitative comparisons (without using the TLL model) between our experiment and previous experiments, where the initial energy distribution function was a double step function [3, 4].

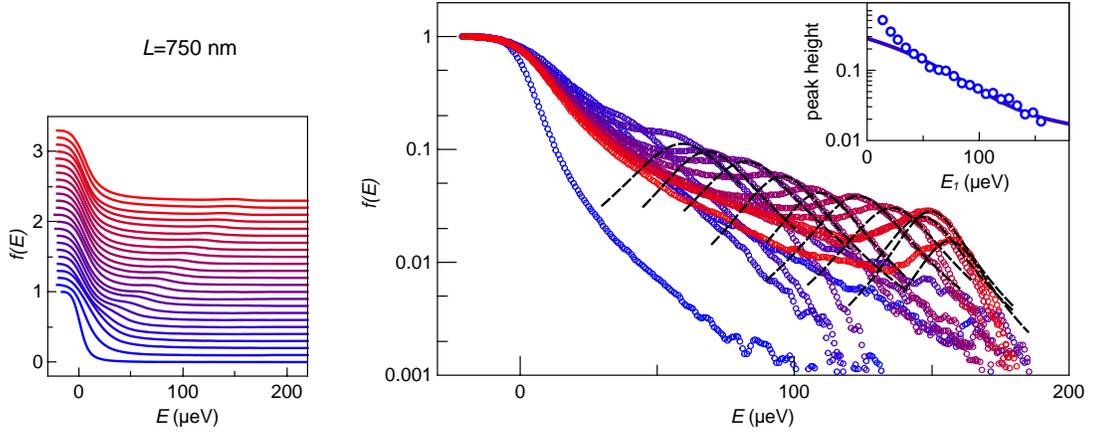

**Supplementary Figure 16 | TLL analysis for additional spectrum R40: 750 nm device.** Left Panel: measured $f(E)$. Each curve, offset for clarity, corresponds to an increment $\delta E \approx 7$ $\mu$eV of the injection energy. Right panel: measured $f(E)$ in semi-log scale (symbols), with Lorentzian fits (black dashed lines). Inset: peak heights (blue circles) extracted from the Lorentzian fits shown in the left panel, plotted in semi-log scale versus $E_1$. The blue line is a fit using the TLL model with dissipation. The fitting parameters are reported in Supplementary Table IV.

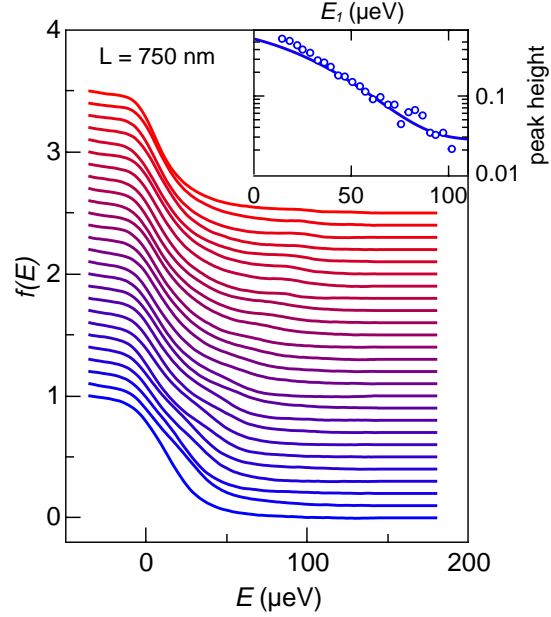

**Supplementary Figure 17 | TLL analysis for additional spectrum R48: 750 nm device.** Measured $f(E)$. Each curve, offset for clarity, corresponds to an increment $\delta E \approx 4$ $\mu$eV of the injection energy. Inset: peak heights $f(E_1)$ (blue circles) plotted in semi-log scale versus $E_1$. The blue line is a fit using the TLL model with dissipation. The fitting parameters are reported in Supplementary Table IV.

A possible quantifier for relaxation is to compare the relative decrease of the quasiparticle peak in our experiment with the relative decrease in the step widths in refs. [3, 4] (as used explicitly in ref. [3]), as both are adimensional quantities that are shown to decrease with propagation length and energy. By doing so, we obtain the following numbers:

- ref. [3]:

* for $L = 0.5$ $\mu$m and $eV_{dc} = 60$ $\mu$eV, the step width in the double step distribution function is reduced by 25 %. In comparison, for a similar length and injection energy $E_1$, the quasiparticle peak amplitude is suppressed by 60 % in our experiment.

* for $L = 0.5$ $\mu$m and $eV_{dc} = 180$ $\mu$eV, the energy spacing of the double step energy distribution function is reduced

| spectrum # | device | length (μm) | $v_2$ (km/s) | $v_\rho$ (km/s) | $v_\sigma$ (km/s) | $\alpha$ | $\theta$ ($\times\pi$) | $\gamma_0$ | $E_0$ (μeV) | $E_\gamma$ (μeV) |
|---|---|---|---|---|---|---|---|---|---|---|
| **S1** – R72 | 480 nm (res. A) | 0.48 | 48 | 89 | 35 | 1.6 | 0.16 | 0.43 | 403 | 142 |
| **S2** – R106 | 480 nm (res. B) | 0.48 | 48 | 92 | 42 | 1.8 | 0.11 | 0.43 | 535 | 147 |
| R76 | 480 nm (long path) | 1.29 | 48 | 94 | 41 | 1.8 | 0.12 | 0.43 | 184 | 55 |
| R84 | 480 nm (separated ECs) | 0.48 | 48 | 87 | 47 | 1.8 | 0.04 | 0.43 | 712 | 139 |
| R99 | 480 nm | 0.48 | 48 | 103 | 31 | 1.8 | 0.16 | 0.43 | 309 | 164 |
| **S3** – R21 | 750 nm (cooldown 1) | 0.75 | 38 | 101 | 17 | 2.1 | 0.17 | 0.13 | 85 | 342 |
| **S4** – R24 | 750 nm (cooldown 2) | 0.75 | 38 | 118 | 18 | 2.6 | 0.15 | 0.12 | 95 | 452 |
| **S5** – R20 | 750 nm | 0.75 | 24 | 62 | 13 | 2.1 | 0.16 | 0.13 | 69 | 209 |
| **S6** – R39 | 750 nm (separated ECs) | 0.75 | 38 | 82 | 36 | 2.1 | 0.07 | 0.13 | 278 | 277 |
| R16 | 750 nm | 0.75 | 23 | 62 | 13 | 2.25 | 0.15 | 0.13 | 72 | 209 |
| R40 | 750 nm | 0.75 | 34 | 97 | 25 | 2.6 | 0.115 | 0.13 | 147 | 329 |
| R48 | 750 nm | 0.75 | 29 | 79 | 17 | 2.4 | 0.14 | 0.13 | 95 | 269 |

**Supplementary Table 4 | TLL fits parameters.** Lowest Fermi velocity $v_2$, charge and dipole plasmon velocities $v_\rho$ and $v_\sigma$, Fermi velocities ratio $\alpha$, effective inter-EC coupling $\theta$, friction coefficient $\gamma_0$, revival energy $E_0$ and exponential decay characteristic energy $E_\gamma$ extracted from the TLL fits.

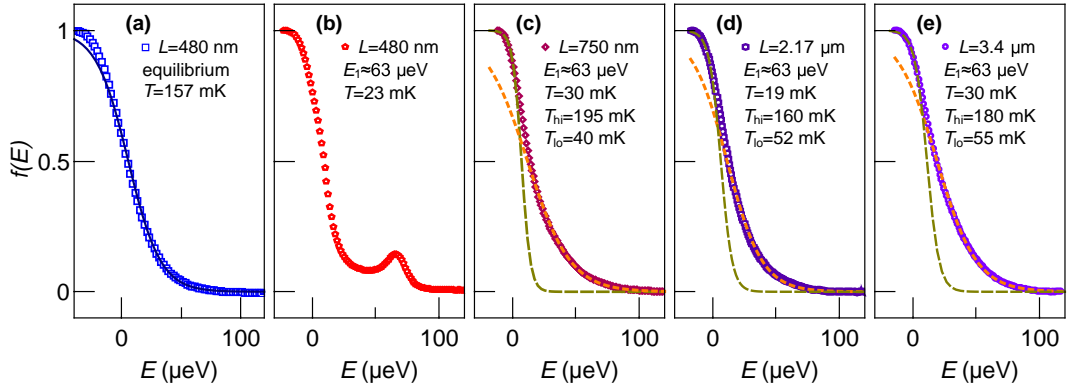

**Supplementary Figure 19 | Length dependence. a,** measured distribution function at equilibrium, for a base temperature of 157 mK (blue squares). The dark blue line is a Fermi function fit of the data with a fixed temperature of 157 mK. **b-e,** measured $f(E)$ when injecting quasiparticles at $E_1 \approx 63$ μeV, with corresponding Fermi function fits (same data as main text Fig. 5). The temperature $T$ indicates the base electron temperature at which each experiment is performed, while $T_{\text{hi}}$ (resp. $T_{\text{lo}}$) indicates the temperature of the Fermi function fit at high (resp. low) energy.

by 50 %. For similar length and injection energy, the quasiparticle peak amplitude is suppressed by 92 % in our

experiment.

- ref. [4]:

* for $L = 0.8$ $\mu$m and $eV_{dc} = 36$ $\mu$eV, the energy spacing of the double step energy distribution function is reduced by 5 %. For similar length and injection energy, the quasiparticle peak amplitude is suppressed by 88 % in our experiment.

These numbers can be put in perspective with the extracted values of the plasmon velocity (the relaxation rate decreases with the velocity): $v^* = 27$ km/s in ref. [3], $v^* = 87$ km/s in ref. [4], and $v^* = 20 - 77$ km/s in our results. Using this quantifier, it thus appears that the quasiparticle peak indeed relaxes "faster", since, for comparable lengths and energies (provided this energy comparison is valid), the relative amplitude suppression is systematically larger than the relative decrease in the double step width. This is emphasized by the fact that the values of the plasmon velocity are similar.

## SUPPLEMENTARY NOTE 8: REFINED TLL MODEL INCLUDING DISSIPATION

We consider the edge channels of a quantum Hall bar at filling factor $\nu = 2$, assuming a short range capacitive coupling between them and energy dissipation due to the external environment. According to the experimental observations no electron tunneling occurs between the edges.

The present experiment involves the injection of an electronic wave-packet with Lorentzian profile in energy and its detection after a given length. In order to describe this situation one can proceed as in Refs. [5–7], where the system is divided into a non-interacting injection region, an interacting propagating region and a non-interacting region of detection.

According to this one can describe the interacting region in terms of the conventional Wen's hydrodynamical approach [8] with Hamiltonian density ($\hbar = 1$)

$$\mathcal{H} = \frac{v_1}{4\pi} (\partial_x \phi_1)^2 + \frac{v_2}{4\pi} (\partial_x \phi_2)^2 + \frac{u}{2\pi} \partial_x \phi_1 \partial_x \phi_2 \qquad (3)$$

where $\phi_1$ and $\phi_2$ are bosonic fields related to the edge particle density through the condition

$$\rho_{1,2} = \frac{1}{2\pi} \partial_x \varphi_{1,2}, \qquad (4)$$

$v_1$ and $v_2$ are the bare propagation velocities of the two channels and $u$ the intensity of their coupling. Without loss of generality in the following we will indicate $v_2 = v$ and $v_1 = \alpha v$, with $\alpha > 1$.

Due to the inter-edge interaction the bosonic fields $\phi_1$ and $\phi_2$ are no longer eigenstates of the Hamiltonian and the system is diagonalized in terms of a charged and a dipole mode, indicated respectively with $\phi_\rho$ and $\phi_\sigma$ with associated eigenvelocities $v_\rho$ and $v_\sigma$ given by

$$v_{\rho,\sigma} = v f_{\rho,\sigma}(\alpha, \theta) \qquad (5)$$

with

$$f_{\rho,\sigma}(\alpha, \theta) = \left(\frac{\alpha+1}{2}\right) \pm \frac{1}{\cos(2\theta)} \left(\frac{\alpha-1}{2}\right) \qquad (6)$$

and

$$\tan(2\theta) = \frac{2u}{v(\alpha-1)}. \qquad (7)$$

It is worth to note that the stability condition of the model imposes the constraint [9]

$$\theta < \frac{1}{2} \arccos\left(\frac{\alpha-1}{\alpha+1}\right) < \frac{\pi}{4}. \qquad (8)$$

The equations of motion, expressed in Fourier transform with respect to time, become

$$(-i\omega + v_\eta \partial_x) \tilde{\phi}_\eta(x, \omega) = 0 \qquad \eta = \rho, \sigma. \qquad (9)$$

Various experiments [4, 10] suggest a relevant role played by energy dissipation in the transport along quantum Hall edge channels. The simplest way to include this effect in the model is by adding a frequency dependent energy

loss rate $\gamma(\omega)$ (assumed here equal for both channels for sake of simplicity) at the level of the equations of motion in the interacting region (see Eq. (9)). According to this, they read

$$[-i\omega + \gamma(\omega) + v_\eta \partial_x] \tilde{\phi}_\eta(x,\omega) = 0 \qquad \eta = \rho, \sigma. \tag{10}$$

In the following we will focus on a linear dependence [9]

$$\gamma(\omega) = \gamma_0 \omega \tag{11}$$

with $\gamma_0$ adimensional friction coefficient, even if more involved functional dependences can be considered [10].

The solution of the equations of motion in Eq. (10) is then given by

$$\tilde{\phi}_\eta(x,\omega) = e^{i\frac{\Gamma\omega}{v_\eta}x}\tilde{\phi}_\eta(0,\omega) \qquad \eta = \rho, \sigma \tag{12}$$

with

$$\Gamma = 1 + i\gamma_0. \tag{13}$$

The initial conditions

$$\begin{aligned}\tilde{\phi}_\rho(0,\omega) &= \cos\theta \tilde{\phi}_1(0,\omega) + \sin\theta \tilde{\phi}_2(0,\omega) \\ \tilde{\phi}_\sigma(0,\omega) &= -\sin\theta \tilde{\phi}_1(0,\omega) + \cos\theta \tilde{\phi}_2(0,\omega)\end{aligned} \tag{14}$$

fix the (possibly frequency dependent) amplitudes at the point of injection $x = 0$.

### Scattering matrix approach

Proceeding as in Ref. [7] we obtain the edge-magnetoplasmon scattering matrix connecting the incoming (injected) and the outgoing (detected) bosonic fields through the relation

$$\begin{pmatrix}\tilde{\phi}_1(L,\omega) \\ \tilde{\phi}_2(L,\omega)\end{pmatrix} = \hat{S}(L,\omega) \begin{pmatrix}\tilde{\phi}_1(0,\omega) \\ \tilde{\phi}_2(0,\omega)\end{pmatrix}, \tag{15}$$

with

$$\hat{S} = \begin{pmatrix}\cos^2\theta e^{i\omega\Gamma\tau_\rho} + \sin^2\theta e^{i\omega\Gamma\tau_\sigma} & \sin\theta\cos\theta\left(e^{i\omega\Gamma\tau_\rho} - e^{i\omega\Gamma\tau_\sigma}\right) \\ \sin\theta\cos\theta\left(e^{i\omega\Gamma\tau_\rho} - e^{i\omega\Gamma\tau_\sigma}\right) & \sin^2\theta e^{i\omega\Gamma\tau_\rho} + \cos^2\theta e^{i\omega\Gamma\tau_\sigma}\end{pmatrix}. \tag{16}$$

In the above equation we have introduced the short-hand notation $\tau_\alpha = L/v_\alpha$ ($\alpha = \rho, \sigma$) for the times of flight associate to the eigenmodes in the interacting region.

In the following we will focus only on the top left entry of the scattering matrix in Eq. (16), namely

$$t(\omega) = \cos^2\theta e^{i\omega\Gamma\tau_\rho} + \sin^2\theta e^{i\omega\Gamma\tau_\sigma} \tag{17}$$

which represents the amplitude probability for the edge-magnetoplasmon to be transmitted along the first channel (injection channel).

### Elastic scattering amplitude

As shown in Ref. [11], assuming a narrow enough (ideally $\delta$-like) injected wave-packet in energy, the evolution of the height of the wave-packet as a function of the injection energy $\xi$ (referred to as $E_1$ in the main text) is given by

$$\mathcal{V}(\xi) = \frac{|\mathcal{Z}(\xi)|^2}{|\mathcal{Z}(0)|^2} \tag{18}$$

with

$$\mathcal{Z}(\xi) = \int_{-\infty}^{+\infty} d\tau e^{i\xi\tau} \exp\left\{\int_0^{+\infty} \frac{d\omega}{\omega}\left[t(\omega)e^{-i\omega\tau} - 1\right]e^{-\omega/\omega_c}\right\} \tag{19}$$

the elastic scattering amplitude, where we introduced a converging factor $\omega_c$ (greatest energy scale in the systems).

By replacing the expression for $t(\omega)$ in Eq. (17) one obtains

$$\mathcal{V}(\xi) = e^{-2\frac{\gamma_0}{f_\rho}\frac{\xi}{\xi_0}} \left| {}_1F_1\left[p_+, 1; -\gamma_0 \frac{\xi}{\xi_0}\left(\frac{1}{f_\sigma} - \frac{1}{f_\rho}\right) + i\frac{\xi}{\xi_0}\left(\frac{1}{f_\sigma} - \frac{1}{f_\rho}\right)\right] \right|^2 \Theta(\xi) \tag{20}$$

with

$$\xi_0 = \frac{\hbar v}{L} \tag{21}$$

and where ${}_1F_1[a, b; z]$ is the Kummer confluent hypergeometric function. This quantity depends on four free parameters, namely $\alpha$, $\theta$, $\gamma_0$ and $\xi_0$, which need to be fixed in order to fit the experimental data.

It is worth noticing that in the "strongly interacting" limit ($\theta = \pi/4$) and in absence of dissipation ($\gamma_0 = 0$) the above expression reduces to

$$\mathcal{V}_{strong}(\xi) = \left| J_0\left(\frac{\xi}{\xi^*}\right) \right|^2 \Theta(\xi) \tag{22}$$

with $J_0$ zero-th order Bessel function and where we introduced the parameter

$$\xi^* = 2\left(\frac{L}{\hbar v_\sigma} - \frac{L}{\hbar v_\rho}\right)^{-1}. \tag{23}$$

This expression is the one considered in Ref. [6].

Notice that the value of $\xi_0$ sets the position of the revival. Moreover, from this value and from the knowledge of the length $L$ of the interacting region reported in the experimental paper it is possible to extract the value of the bare velocity $v$.

[1] C. Rössler, S. Burkhard, T. Krähenmann, M. Röösli, P. Märki, J. Basset, T. Ihn, K. Ensslin, C. Reichl, and W. Wegscheider, Physical Review B **90**, 081302 (2014).
[2] I. P. Levkivskyi and E. V. Sukhorukov, Physical Review B **85**, 075309 (2012).
[3] K. Itoh, R. Nakazawa, T. Ota, M. Hashisaka, K. Muraki, and T. Fujisawa, Physical Review Letters **120**, 197701 (2018).
[4] H. le Sueur, C. Altimiras, U. Gennser, A. Cavanna, D. Mailly, and F. Pierre, Physical Review Letters **105**, 056803 (2010).
[5] P. Degiovanni, C. Grenier, G. Fève, C. Altimiras, H. le Sueur, and F. Pierre, Physical Review B **81**, 121302 (2010).
[6] D. Ferraro, B. Roussel, C. Cabart, E. Thibierge, G. Fève, C. Grenier, and P. Degiovanni, Physical Review Letters **113**, 166403 (2014).
[7] D. Ferraro and E. Sukhorukov, SciPost Physics **3**, 014 (2017).
[8] X.-G. Wen, Adv. Phys. **44**, 405 (1995).
[9] A. Braggio, D. Ferraro, M. Carrega, N. Magnoli, and M. Sassetti, New J. Phys. **14**, 093032 (2012).
[10] E. Bocquillon, V. Freulon, J.-M. Berroir, P. Degiovanni, B. Plaçais, A. Cavanna, Y. Jin, and G. Fève, Nature Communications **4**, 1839 (2013).
[11] P. Degiovanni, C. Grenier, and G. Fève, Physical Review B **80**, 241307 (2009).




\begin{thebibliography}{46}
\expandafter\ifx\csname natexlab\endcsname\relax\def\natexlab#1{#1}\fi
\expandafter\ifx\csname url\endcsname\relax
 \def\url#1{\texttt{#1}}\fi
\expandafter\ifx\csname urlprefix\endcsname\relax\def\urlprefix{URL }\fi

\bibitem[{Bocquillon \textit{et~al.}(2014)Bocquillon et~al.}]{Bocquillon2014}
Bocquillon, E. et~al.
\newblock {Electron quantum optics in ballistic chiral conductors}.
\newblock \textit{Annalen der Physik} \textbf{526} (2014).

\bibitem[{B{\"{a}}uerle \textit{et~al.}(2018)B{\"{a}}uerle
 et~al.}]{Bauerle2018}
B{\"{a}}uerle, C. et~al.
\newblock {Coherent control of single electrons: a review of current progress}.
\newblock \textit{Reports on Progress in Physics} \textbf{81}, 056503 (2018).

\bibitem[{Ji \textit{et~al.}(2003)Ji et~al.}]{Ji2003}
Ji, Y. et~al.
\newblock {An electronic Mach-Zehnder interferometer}.
\newblock \textit{Nature} \textbf{422}, 415--418 (2003).

\bibitem[{Bocquillon \textit{et~al.}(2013{\natexlab{a}})Bocquillon
 et~al.}]{Bocquillon2013a}
Bocquillon, E. et~al.
\newblock {Coherence and Indistinguishability of Single Electrons Emitted by
 Independent Sources}.
\newblock \textit{Science} \textbf{339}, 1054--1057 (2013{\natexlab{a}}).

\bibitem[{Neder \textit{et~al.}(2006)Neder, Heiblum, Levinson, Mahalu \&
 Umansky}]{Neder2006}
Neder, I., Heiblum, M., Levinson, Y., Mahalu, D. \& Umansky, V.
\newblock {Unexpected Behavior in a Two-Path Electron Interferometer}.
\newblock \textit{Physical Review Letters} \textbf{96}, 016804 (2006).

\bibitem[{Roulleau \textit{et~al.}(2007)Roulleau et~al.}]{Roulleau2007}
Roulleau, P. et~al.
\newblock {Finite bias visibility of the electronic Mach-Zehnder
 interferometer}.
\newblock \textit{Physical Review B} \textbf{76}, 161309 (2007).

\bibitem[{Litvin \textit{et~al.}(2008)Litvin, Helzel, Tranitz, Wegscheider \&
 Strunk}]{Litvin2008}
Litvin, L.~V., Helzel, A., Tranitz, H.-P., Wegscheider, W. \& Strunk, C.
\newblock {Edge-channel interference controlled by Landau level filling}.
\newblock \textit{Physical Review B} \textbf{78}, 075303 (2008).

\bibitem[{Roulleau \textit{et~al.}(2008{\natexlab{a}})Roulleau
 et~al.}]{Roulleau2008}
Roulleau, P. et~al.
\newblock {Noise Dephasing in Edge States of the Integer Quantum Hall Regime}.
\newblock \textit{Physical Review Letters} \textbf{101}, 186803
 (2008{\natexlab{a}}).

\bibitem[{Roulleau \textit{et~al.}(2008{\natexlab{b}})Roulleau
 et~al.}]{Roulleau2008a}
Roulleau, P. et~al.
\newblock {Direct Measurement of the Coherence Length of Edge States in the
 Integer Quantum Hall Regime}.
\newblock \textit{Physical Review Letters} \textbf{100}, 126802
 (2008{\natexlab{b}}).

\bibitem[{Altimiras \textit{et~al.}(2009)Altimiras et~al.}]{Altimiras2009}
Altimiras, C. et~al.
\newblock {Non-equilibrium edge-channel spectroscopy in the integer quantum
 Hall regime}.
\newblock \textit{Nature Physics} \textbf{6}, 34--39 (2009).

\bibitem[{le~Sueur \textit{et~al.}(2010)le~Sueur et~al.}]{leSueur2010}
le~Sueur, H. et~al.
\newblock {Energy Relaxation in the Integer Quantum Hall Regime}.
\newblock \textit{Physical Review Letters} \textbf{105}, 056803 (2010).

\bibitem[{Sukhorukov \& Cheianov(2007)}]{Sukhorukov2007}
Sukhorukov, E. \& Cheianov, V.
\newblock {Resonant Dephasing in the Electronic Mach-Zehnder Interferometer}.
\newblock \textit{Physical Review Letters} \textbf{99}, 156801 (2007).

\bibitem[{Levkivskyi \& Sukhorukov(2008)}]{Levkivskyi2008}
Levkivskyi, I. \& Sukhorukov, E.
\newblock {Dephasing in the electronic Mach-Zehnder interferometer at filling
 factor $\nu$=2}.
\newblock \textit{Physical Review B} \textbf{78}, 045322 (2008).

\bibitem[{Degiovanni \textit{et~al.}(2009)Degiovanni, Grenier \&
 F{\`{e}}ve}]{Degiovanni2009}
Degiovanni, P., Grenier, C. \& F{\`{e}}ve, G.
\newblock {Decoherence and relaxation of single-electron excitations in quantum
 Hall edge channels}.
\newblock \textit{Physical Review B} \textbf{80}, 241307 (2009).

\bibitem[{Degiovanni \textit{et~al.}(2010)Degiovanni et~al.}]{Degiovanni2010}
Degiovanni, P. et~al.
\newblock {Plasmon scattering approach to energy exchange and high-frequency
 noise in $\nu$ = 2 quantum Hall edge channels}.
\newblock \textit{Physical Review B} \textbf{81}, 121302 (2010).

\bibitem[{Ferraro \textit{et~al.}(2014)Ferraro et~al.}]{Ferraro2014}
Ferraro, D. et~al.
\newblock {Real-Time Decoherence of Landau and Levitov Quasiparticles in
 Quantum Hall Edge Channels}.
\newblock \textit{Physical Review Letters} \textbf{113}, 166403 (2014).

\bibitem[{Freulon \textit{et~al.}(2015)Freulon et~al.}]{Freulon2015}
Freulon, V. et~al.
\newblock {Hong-Ou-Mandel experiment for temporal investigation of
 single-electron fractionalization}.
\newblock \textit{Nature Communications} \textbf{6}, 6854 (2015).

\bibitem[{Marguerite \textit{et~al.}(2016)Marguerite et~al.}]{Marguerite2016}
Marguerite, A. et~al.
\newblock {Decoherence and relaxation of a single electron in a one-dimensional
 conductor}.
\newblock \textit{Physical Review B} \textbf{94}, 115311 (2016).

\bibitem[{Marguerite \textit{et~al.}(2017)Marguerite et~al.}]{Marguerite2017}
Marguerite, A. et~al.
\newblock {Two-particle interferometry in quantum Hall edge channels}.
\newblock \textit{Physica Status Solidi (b)} \textbf{254}, 1600618 (2017).

\bibitem[{Gurman \textit{et~al.}(2016)Gurman, Sabo, Heiblum, Umansky \&
 Mahalu}]{Gurman2016}
Gurman, I., Sabo, R., Heiblum, M., Umansky, V. \& Mahalu, D.
\newblock {Dephasing of an electronic two-path interferometer}.
\newblock \textit{Physical Review B} \textbf{93}, 121412 (2016).

\bibitem[{Tewari \textit{et~al.}(2016)Tewari et~al.}]{Tewari2016}
Tewari, S. et~al.
\newblock {Robust quantum coherence above the Fermi sea}.
\newblock \textit{Physical Review B} \textbf{93}, 035420 (2016).

\bibitem[{Kr{\"{a}}henmann \textit{et~al.}(2019)Kr{\"{a}}henmann
 et~al.}]{Krahenmann2019}
Kr{\"{a}}henmann, T. et~al.
\newblock {Auger-spectroscopy in quantum Hall edge channels and the missing
 energy problem}.
\newblock \textit{Nature Communications} \textbf{10}, 3915 (2019).

\bibitem[{Fischer \textit{et~al.}(2019)Fischer, Park, Meir \&
 Gefen}]{Fischer2019}
Fischer, S.~G., Park, J., Meir, Y. \& Gefen, Y.
\newblock {Interaction-induced charge transfer in a mesoscopic electron
 spectrometer}.
\newblock \textit{Physical Review B} \textbf{100}, 195411 (2019).

\bibitem[{Lunde \& Nigg(2016)}]{Lunde2016}
Lunde, A.~M. \& Nigg, S.~E.
\newblock {Statistical theory of relaxation of high-energy electrons in quantum
 Hall edge states}.
\newblock \textit{Physical Review B} \textbf{94}, 045409 (2016).

\bibitem[{Takei \textit{et~al.}(2010)Takei, Milletar{\`{i}} \&
 Rosenow}]{Takei2010}
Takei, S., Milletar{\`{i}}, M. \& Rosenow, B.
\newblock {Nonequilibrium electron spectroscopy of Luttinger liquids}.
\newblock \textit{Physical Review B} \textbf{82}, 041306 (2010).

\bibitem[{Altimiras \textit{et~al.}(2010)Altimiras et~al.}]{Altimiras2010}
Altimiras, C. et~al.
\newblock {Tuning Energy Relaxation along Quantum Hall Channels}.
\newblock \textit{Physical Review Letters} \textbf{105}, 226804 (2010).

\bibitem[{Itoh \textit{et~al.}(2018)Itoh et~al.}]{Itoh2018}
Itoh, K. et~al.
\newblock {Signatures of a Nonthermal Metastable State in Copropagating Quantum
 Hall Edge Channels}.
\newblock \textit{Physical Review Letters} \textbf{120}, 197701 (2018).

\bibitem[{R{\"{o}}ssler \textit{et~al.}(2014)R{\"{o}}ssler
 et~al.}]{Rossler2014}
R{\"{o}}ssler, C. et~al.
\newblock {Spectroscopy of equilibrium and nonequilibrium charge transfer in
 semiconductor quantum structures}.
\newblock \textit{Physical Review B} \textbf{90}, 081302 (2014).

\bibitem[{Ota \textit{et~al.}(2019)Ota, Akiyama, Hashisaka, Muraki \&
 Fujisawa}]{Ota2019}
Ota, T., Akiyama, S., Hashisaka, M., Muraki, K. \& Fujisawa, T.
\newblock {Spectroscopic study on hot-electron transport in a quantum Hall edge
 channel}.
\newblock \textit{Physical Review B} \textbf{99}, 085310 (2019).

\bibitem[{Gabelli \& Reulet(2013)}]{Gabelli2013}
Gabelli, J. \& Reulet, B.
\newblock {Shaping a time-dependent excitation to minimize the shot noise in a
 tunnel junction}.
\newblock \textit{Physical Review B} \textbf{87}, 075403 (2013).

\bibitem[{Grenier \textit{et~al.}(2011)Grenier et~al.}]{Grenier2011b}
Grenier, C.~C. et~al.
\newblock {Single-electron quantum tomography in quantum Hall edge channels}.
\newblock \textit{New Journal of Physics} \textbf{13}, 093007 (2011).

\bibitem[{Jullien \textit{et~al.}(2014)Jullien et~al.}]{Jullien2014}
Jullien, T. et~al.
\newblock {Quantum tomography of an electron}.
\newblock \textit{Nature} \textbf{514}, 603--607 (2014).

\bibitem[{Bisognin \textit{et~al.}(2019)Bisognin et~al.}]{Bisognin2019}
Bisognin, R. et~al.
\newblock {Quantum tomography of electrical currents}.
\newblock \textit{Nature Communications} \textbf{10}, 3379 (2019).

\bibitem[{Fletcher \textit{et~al.}(2019)Fletcher et~al.}]{Fletcher2019}
Fletcher, J.~D. et~al.
\newblock {Continuous-variable tomography of solitary electrons}.
\newblock \textit{Nature Communications} \textbf{10}, 5298 (2019).

\bibitem[{Bocquillon \textit{et~al.}(2013{\natexlab{b}})Bocquillon
 et~al.}]{Bocquillon2013b}
Bocquillon, E. et~al.
\newblock {Separation of neutral and charge modes in one-dimensional chiral
 edge channels}.
\newblock \textit{Nature Communications} \textbf{4}, 1839 (2013{\natexlab{b}}).

\bibitem[{Braggio \textit{et~al.}(2012)Braggio, Ferraro, Carrega, Magnoli \&
 Sassetti}]{Braggio2012}
Braggio, A., Ferraro, D., Carrega, M., Magnoli, N. \& Sassetti, M.
\newblock {Environmental induced renormalization effects in quantum Hall edge
 states due to 1/f noise and dissipation}.
\newblock \textit{New Journal of Physics} \textbf{14}, 093032 (2012).

\bibitem[{Slobodeniuk \textit{et~al.}(2016)Slobodeniuk, Idrisov \&
 Sukhorukov}]{Slobodeniuk2016}
Slobodeniuk, A.~O., Idrisov, E.~G. \& Sukhorukov, E.~V.
\newblock {Relaxation of an electron wave packet at the quantum Hall edge at
 filling factor $\nu$ = 2}.
\newblock \textit{Physical Review B} \textbf{93}, 035421 (2016).

\bibitem[{Goremykina \& Sukhorukov(2018)}]{Goremykina2018}
Goremykina, A.~S. \& Sukhorukov, E.~V.
\newblock {Coherence recovery mechanisms of quantum Hall edge states}.
\newblock \textit{Physical Review B} \textbf{97}, 115418 (2018).

\bibitem[{Hashisaka \textit{et~al.}(2017)Hashisaka, Hiyama, Akiho, Muraki \&
 Fujisawa}]{Hashisaka2017}
Hashisaka, M., Hiyama, N., Akiho, T., Muraki, K. \& Fujisawa, T.
\newblock {Waveform measurement of charge- and spin-density wavepackets in a
 chiral Tomonagaâ€“Luttinger liquid}.
\newblock \textit{Nature Physics} \textbf{13}, 559--562 (2017).

\bibitem[{Inoue \textit{et~al.}(2014)Inoue et~al.}]{Inoue2014}
Inoue, H. et~al.
\newblock {Charge Fractionalization in the Integer Quantum Hall Effect}.
\newblock \textit{Physical Review Letters} \textbf{112}, 166801 (2014).

\bibitem[{Levkivskyi \& Sukhorukov(2012)}]{Levkivskyi2012}
Levkivskyi, I.~P. \& Sukhorukov, E.~V.
\newblock {Energy relaxation at quantum Hall edge}.
\newblock \textit{Physical Review B} \textbf{85}, 075309 (2012).

\bibitem[{Huynh \textit{et~al.}(2012)Huynh et~al.}]{Huynh2012}
Huynh, P.-A. et~al.
\newblock {Quantum Coherence Engineering in the Integer Quantum Hall Regime}.
\newblock \textit{Physical Review Letters} \textbf{108}, 256802 (2012).

\bibitem[{Cabart \textit{et~al.}(2018)Cabart, Roussel, F{\`{e}}ve \&
 Degiovanni}]{Cabart2018}
Cabart, C., Roussel, B., F{\`{e}}ve, G. \& Degiovanni, P.
\newblock {Taming electronic decoherence in one-dimensional chiral ballistic
 quantum conductors}.
\newblock \textit{Physical Review B} \textbf{98}, 155302 (2018).

\bibitem[{Duprez \textit{et~al.}(2019)Duprez et~al.}]{Duprez2019}
Duprez, H. et~al.
\newblock {Macroscopic Electron Quantum Coherence in a Solid-State Circuit}.
\newblock \textit{Physical Review X} \textbf{9}, 021030 (2019).

\bibitem[{Goremykina \textit{et~al.}(2019)Goremykina, Borin \&
 Sukhorukov}]{Goremykina2019}
Goremykina, A., Borin, A. \& Sukhorukov, E.
\newblock {Heat current in a dissipative quantum Hall edge}.
\newblock Preprint at \href{http://arxiv.org/abs/1908.01213}{http://arxiv.org/abs/1908.01213} (2019).

\end{thebibliography}

\end{document}